\documentclass{aa}  
\usepackage{graphicx}
\usepackage{txfonts}
\usepackage{multirow}
\usepackage{color}
\usepackage{multirow}
\usepackage{ulem}

\newcommand{\teff}{$T_{\rm eff}$}

\definecolor{rojo}{rgb}{1,0,0}
\definecolor{blu}{rgb}{0,0,1}
\definecolor{mag}{rgb}{1,0,1}

\newcommand{\kms}{km\,s$^{-1}$}

\begin{document}

\title{High-resolution spectroscopy of the young open cluster M\,39 (NGC\,7092)}

\author{J. Alonso-Santiago\inst{1}\and
	A. Frasca\inst{1}\and 
        G. Catanzaro\inst{1}\and
        A. Bragaglia\inst{2}\and
        L. Magrini\inst{3}\and  
        A. Vallenari\inst{4}\and
        E. Carretta\inst{2}\and
        S. Lucatello\inst{4}  
        }

\institute{
INAF--Osservatorio Astrofisico di Catania, via S. Sofia 78, 95123 Catania, Italy
\and
INAF--Osservatorio di Astrofisica e Scienza dello Spazio, via P. Gobetti 93/3, 40129 Bologna, Italy
\and
INAF--Osservatorio Astrofisico di Arcetri,  Largo  E.  Fermi  5, 50125 Firenze, Italy
\and
INAF--Osservatorio Astronomico di Padova, vicolo dell'Osservatorio 5, 35122 Padova, Italy
}

\date{}

 
  \abstract
  {M\,39 is a nearby young open cluster hardly studied in the last decades. No giant is known among its members and its chemical composition has never been
  studied. In order to investigate it we performed high-resolution spectroscopy of 20 expected cluster members with the HARPS and FIES spectrographs. By combining our
  observations with archival photometry and $Gaia$-DR3 data we searched for evolved members and studied cluster properties such as the radial velocity, 
  extinction and age. For the first time, we provide stellar parameters and chemical abundances for 21 species with atomic numbers up to 56. We have not 
  found any new giant as likely member and notice a negligible reddening along the cluster field, that we place at 300\,pc. We obtain a mean radial velocity 
  for M\,39 of $-$5.5$\pm$0.5\,km\,s$^{-1}$ and an isochrone-fitting age of 430\,$\pm$\,110 Ma, which corresponds to a MSTO mass of around 2.8 M$_{\sun}$. This value is consistent with the Li content and chromospheric activity shown by its members. Based on main-sequence stars
  the cluster exhibits a solar composition, [Fe/H]=$+$0.04$\pm$0.08\,dex, compatible with its Galactocentric location. However, it has a slightly subsolar abundance 
  of Na and an enriched content of neutron-capture elements, specially Ba. In any case, the chemical composition of M\,39 is fully compatible with that shown
  by other open clusters that populate the Galactic thin disc.

  }
  

	
   \titlerunning{High-resolution spectroscopy of M\,39}
   \maketitle
%

\section{Introduction}\label{Sec:intro}
 
The abundance and spatial distribution of the different chemical elements as well as their variation over time, are keys to understanding the origin 
and evolution of our Galaxy. In the era of $Gaia$ and the large spectroscopic surveys, such as for example $Gaia$-ESO \citep{Gilmore22,Randich22}, 
GALAH \citep{GALAH} and APOGEE \citep{APOGEE}, Galactic Archaeology is experiencing an unprecedent advance, significantly improving our knowledge of the
structures and history of the Milky Way. Open clusters (OCs) are among the best tracers of the Galactic thin disc, since they are disseminated throughout it,
covering a wide range of distances and ages. Furthermore, the chemical cluster composition, usually based on the analysis of a large number of members, is more
reliable than that obtained from individual field stars. The best tool for deriving accurate chemical abundances is high-resolution spectroscopy.
However, despite these large surveys and some other that are about to come into operation like WEAVE \citep{WEAVE} and 4MOST \citep{4MOST}, only a small
fraction of OCs (around 10\%) have been observed in this way to date. 
This fraction could be even lower if we consider the recently discovered new OCs \citep{Hunt23} in light of the $Gaia$ third data release \citep[$Gaia$-DR3,][]{DR3}.
Thus, high-resolution spectroscopy, such as the one conducted in this work, is highly 
required to continue studying our Galaxy in depth. 

M\,39 (also known as NGC\,7092) is a sparse and nearby OC in Cygnus. It was first studied by \citet{Trumpler28}, who classified it as a type 1a 
object. That means the cluster hosts only main sequence (MS) stars, i.e. with no late-type giants, not earlier than spectral type A. 
The most relevant works focused on M\,39 date back several decades \citep[see e.g., ][]{Ebbighausen40,Eggen51,Johnson53,Weaver53,Abt73,Abt76,McNamara77,Mohan85,
Manteiga91,Platais94}. The cluster has been mostly studied by combining photographic and photoelectric $UBV$ photometry, spectral types obtained with low-resolution
slit spectroscopy and membership identification based on proper motions derived from photographic plates. These works determined the cluster properties by observing a
small number of stars, a few dozens at best. Nonetheless, \citet{Platais94} provided photometry and proper motions for 7931 stars in the field of M\,39, assigning
spectral types for 511 of them. 

The aforementioned studies agree that the cluster lies at around 300 pc and that is practically unaffected by interstellar extinction.
Among the MS stars the earliest spectral type is found to be B9-A0 \citep[e.g.][]{Abt76}, confirming the pioneering result of \citet{Trumpler28}. As there is no
evolved members it has not been easy to accurately determine the age of the cluster, which appears to be in the 200--500 Ma\footnote{Throughout the text we 
follow the recommendations of the IAU using the symbol $a$ (from the Latin word $annus$) as the unit of time rather than $yr$.} range \citep{Mohan85,Manteiga91}.

Recently, \citet{CG20}, based on the $Gaia$ second data release, placed M\,39 at 311 pc, found no appreciable reddening in the cluster field and estimated an age of about 400 Ma. To date, M\,39 has not yet been observed with high-resolution spectroscopy and its chemical composition remains unknown. 
The main aim of  this paper is to fill this gap. We conducted the present investigation using high-resolution spectra, archival photometry and $Gaia$-DR3 data
with the aim of characterizing the main properties of the cluster. We payed particular attention to the determination of the chemical abundances of a large number
of elements that allow us to perform the proper chemical tagging of our target. The paper is structured as follows. We present our observations in Sect.~\ref{Sec:obs}
and explain the search performed to identify new (evolved) cluster members in Sect.~\ref{Sec:memb}. Then, in Sect.~\ref{Sec:sp} we describe our spectral analysis 
and in Sect.~\ref{Sec:sed} we study the cluster reddening. The investigation of the age is outlined in Sect.~\ref{Sec:CMD}, while the discussion and comparison of our
main results with the literature are presented in Sect.~\ref{Sec:disc}. Finally, we summarise our results and present the conclusions of this work in 
Sect.~\ref{Sec:concl}.

\section{Observations}\label{Sec:obs}       

A fundamental step when studying a stellar cluster is to carefully choose a representative sample of members.
In this work the target selection was performed
before the publication of the $Gaia$-DR2 catalogue. Therefore, as a starting point, we made use of the list of members identified by \citet{Platais94}, the most comprehensive study focused on 
M\,39 available in the literature. Then, we completed our target list taking advantage of the selection made by \citet{CG18}. As noted above, the cluster does not 
host any cool giants and, therefore, our sample consists of MS stars only. For this reason, we payed special attention to the brightest stars located in the upper MS and the turn-off 
point (TO) that can help us to constrain the cluster age. 

\subsection{Spectroscopy}

The observations were conducted mainly in October 2017. High-resolution spectroscopy was taken for our targets by employing two different spectrographs, both located 
at \textit{El Roque de los Muchachos} Observatory (La Palma, Spain). Most of the sources, 17, were observed with the High Accuracy Radial velocity Planet Searcher 
for the Northern hemisphere spectrograph \citep[HARPS-N,][]{Cosentino14} while three other spectra were obtained with the Fibre-fed Echelle Spectrograph 
\citep[FIES,][]{Telting14}. HARPS-N is an {\'e}chelle spectrograph fibre-fed from the Nasmyth B focus of the 3.6-m \textit{Telescopio Nazionale Galileo} (TNG).
It covers the wavelength range between 3900 and 6800\,\AA{} in 68 orders and provides a resolving power of $R$=115\,000. As regards FIES, it is mounted at the 
the 2.5-m Nordic Optical Telescope (NOT). It covers the spectral range 3700-9000\,\AA{} in 81 orders, being able to work at three different resolutions. We used 
the high-resolution mode, which provides a $R$=67\,000. Addtionally, three other spectra were taken with HARPS-N in 2020. 

The spectra were acquired with exposures times ($t_{\rm exp}$) ranging from 300 to 7200 s, depending on the star brightness and sky conditions. To reduce the contamination 
of cosmic rays, we avoid taking exposures longer than 1800 s. A standard reduction of both sets of spectra was performed using the corresponding instrument 
pipeline. In total we collected 24 spectra for 20 different stars, which are listed in Table~\ref{Tab:obs_log}. Throughout this paper we follow the numbering 
assigned by \citet{Platais94} to identify our targets. The star UCAC4~683-104396, a bona fide member according to \citet{CG18} observed in our last run, 
is out of the field surveyed by \citet{Platais94}. For this reason we use ``ucac'' as a shorthand designation for this star.

\begin{table}
\caption{Observation log.}
\begin{center}
\begin{tabular}{lcccc}   
\hline\hline
Star$^{a}$ &  Name &  Date   &    $t_{\rm exp}$ (s) & S/N$^b$ \\ 
\hline
\multicolumn{5}{c}{HARPS-N}\\
\hline
0305 & \object{TIC~64397736}	 & 2017-10-03  & 7200  &  46 \\ 
0305 & \object{TIC~64397736}	 & 2020-08-02  & 6300  &  53 \\ 
3004 & \object{BD+47\,3440}	 & 2017-10-04  & 3600  &  40 \\ 
3061 & \object{TYC 3594-834-1}   & 2017-10-29  & 7200  &  92 \\ 
3288 & \object{TIC~64983141}	 & 2017-12-05  & 7200  &  76 \\ 
3311 & \object{BD+47\,3442}	 & 2017-10-29  & 3600  & 105 \\ 
3423 & \object{TYC 3594-678-1}   & 2017-10-04  & 7200  &  92 \\ 
3781 & \object{HD~205116}	 & 2017-10-29  &  600  & 126 \\ 
3814 & \object{HD~205117}	 & 2017-10-29  & 1200  & 118 \\ 
4265 & \object{HD~205198}	 & 2017-10-03  & 1200  & 155 \\ 
4322 & \object{TIC 417208749}	 & 2017-10-28  & 7200  &  35 \\ 
4438 & \object{TYC 3598-511-1}   & 2017-10-28  & 7200  &  47 \\ 
4673 & \object{BD+47\,3462}	 & 2017-10-29  & 1800  &  71 \\ 
5045 & \object{HD~205331}	 & 2017-10-29  & 1200  & 174 \\ 
5609 & \object{TYC~3599-2886-1}  & 2017-10-04  & 2400  &  55 \\ 
5609 & \object{TYC~3599-2886-1}  & 2017-10-28  & 1200  &  40 \\ 
5729 & \object{BD+48\,3423}	 & 2017-10-03  & 3600  & 161 \\ 
6791 & \object{BD+47\,3480}	 & 2017-10-29  & 3600  & 131 \\ 
ucac & \scriptsize{\object{UCAC4~683-104396}} &  2020-09-08  &  4200 & 29 \\ 
ucac & \scriptsize{\object{UCAC4~683-104396}} &  2020-11-24  &  2100 & 38 \\ 
\hline
\multicolumn{5}{c}{FIES}\\
\hline
2451 & \object{HD~204917}	 & 2017-10-23  &  900 &  65 \\ 
4294 & \object{HD~205210}	 & 2017-10-23  & 1200 & 100 \\ 
7140 & \object{BD+47\,3485}	 & 2017-10-22  &  300 &  85 \\ 
7140 & \object{BD+47\,3485}	 & 2017-10-24  &  300 &  60 \\ 
\hline  
\end{tabular}
\begin{list}{}{}
\item[$^a$] Designation from \citet{Platais94} 
\item[$^b$] Signal-to-noise ratio per pixel at 6500\,\AA.
\end{list}
\label{Tab:obs_log}
\end{center}
\end{table}

\subsection{Archival data}       
   
To complement our spectroscopic observations we resorted to archival data provided by some all-sky surveys. This information will be very useful later as it 
will allow us to study both the cluster members and the cluster itself. On the one hand, we used photometric data in the optical range from the APASS catalogue
\citep{APASS} and in the near-infrared wavelengths from the 2MASS \citep{2MASS} and WISE \citep{WISE} catalogues. On the other hand, we took advantage of the 
last data release, DR3, of the $Gaia$ mission \citep{DR3}. Tables~\ref{tab_astrom} and \ref{tab_fotom} report the $Gaia$-DR3 source identifier as well as the photometric and astrometric
data for the stars observed in this work at the end of the paper.

\section{Cluster membership}\label{Sec:memb}

The disentangling of cluster members from field stars is not a trivial task. However, to succeed in the characterisation of the cluster, a correct identification
of its members is essential. The arrival of a large amount of high-precision photometric and, specially, astrometric data from the $Gaia$ mission, has been of 
great help in simplifying this task. 

\citet{Platais94} observed about 8000 stars in the field of M\,39 and identified 511 likely members among them. His analysis was based on $UBV$ photometry and proper
motions measured on photographic plates. As above mentioned, we selected our targets from this list. Specifically, we took 19 high-probability members, that is, with an 
assigned membership probability, $P\ge$0.7, among the brightest.

\citet{CG18} taking advantage of the $Gaia$-DR2 identified 198 members, out of which 154 bona fide members (i.e. with $P\ge$0.7) were found. Among our targets, 14 are also 
high-probability members according to them, while the remaining five stars, namely 3004, 3423, 4294, 5045, and 5609, are not considered by them as cluster members.
In a second run, we observed the star ucac, a likely member reported in \citet{CG18}, out of the field studied by \citet{Platais94}. 

With the intention of finding new members, and specially some giant that allow us to improve the estimation of the cluster age when performing the isochrone 
fitting, we carried out a search around the nominal centre of the cluster. We took all the sources from the $Gaia$-DR3 catalogue brighter than $G$=18 within a
radius of 250\arcmin around it. In total we collected $\approx$1\,250\,000 objects with a complete set of astrometric (i. e. $\mu_{\alpha*}$, $\mu_\delta$, $\varpi$) and 
photometric ($G$, $G_{\textrm{BP}}$, $G_{\textrm{RP}}$) parameters.  

\citet{CG18} placed M\,39 in the astrometric space at ($\mu_{\alpha*}$,$\mu_\delta$,$\varpi$)\,=\,($-$7.477, $-$19.738, 3.337)\,$\pm$\,(0.381, 0.393, 0.090)
(mas, mas\,a$^{-1}$, mas\,a$^{-1}$). When updating the $Gaia$ parameters from DR2 to DR3 for the members reported in that paper we noticed a non negligible 
dispersion in the astrometric data, but not in the photometric values (see Figs.~\ref{astrom_dr2dr3} and \ref{photom_dr2dr3}). For this reason we decided to 
recalculate the cluster centre adopting the average DR3 values of their bona fide members. Thus, we now found the cluster at 
($\mu_{\alpha*}$,$\mu_\delta$,$\varpi$)\,=\,($-$7.459, $-$19.839, 3.367)\,$\pm$\,(0.345, 0.419, 0.088), where the errors correspond to the standard deviation.
Once we determined the cluster centre in the astrometric space, we adopted a radius around it of 3$\sigma$, 4$\sigma$ and 5$\sigma$ as a membership 
criterion to define three classes of members: type I (high-probability), type II (medium-probability) and type III (marginal membership), respectively. In this
way, we identified 196, 48 and 27 likely members within these three radii, finding 147, 5 and 0 bona fide members from \citet{CG18}, respectively.

\begin{figure}[!h]
\begin{center}

\includegraphics[width=7cm]{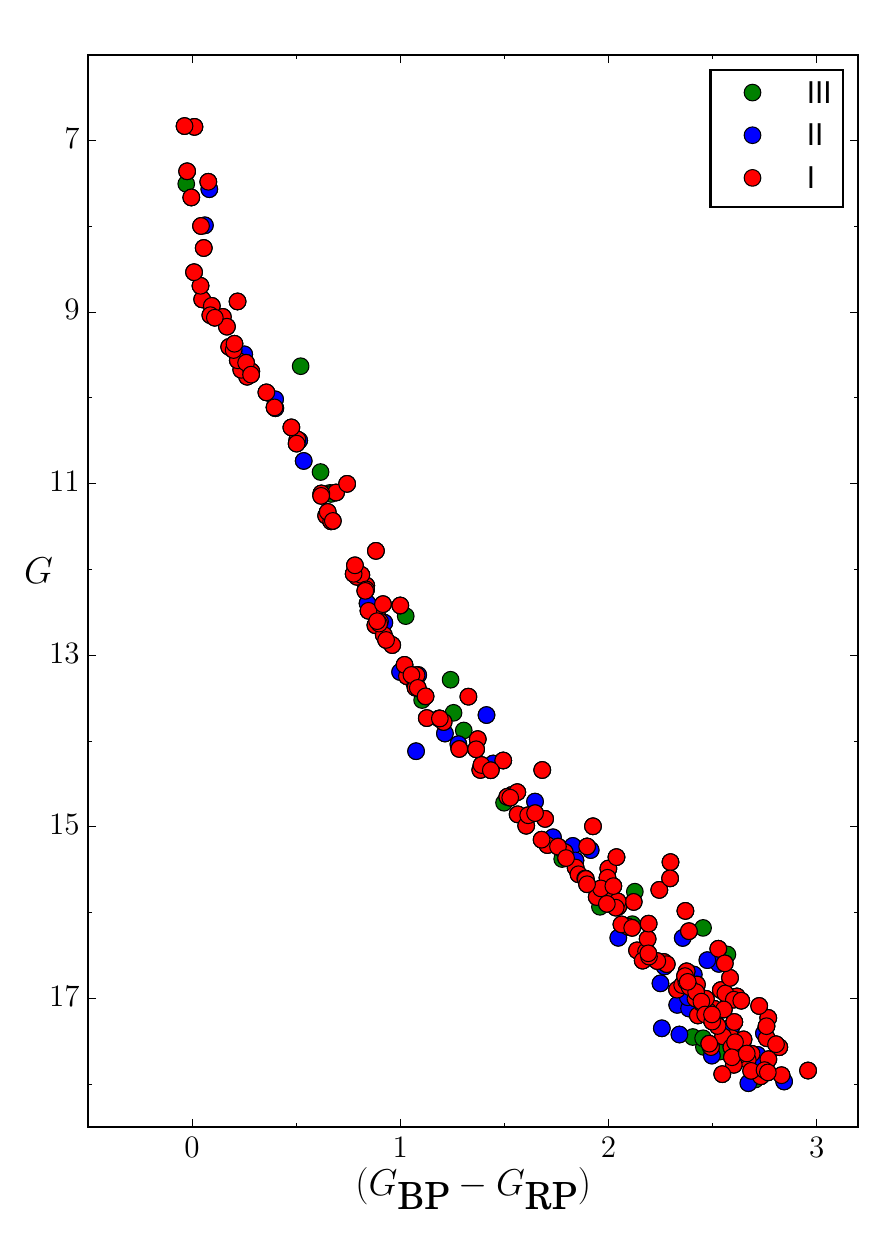}	
\caption{$Gaia$ colour-magnitude diagram of M\,39 displaying the members identified in this work. The colour indicates the membership class, as defined in the text.}

\label{fig_mp2}
\end{center}
\end{figure}

In a second step we refined our selection of members by means of photometry. We plotted the ($G_{\textrm{BP}}-G_{\textrm{RP}}$)/$G$ diagram and rejected those 
objects (six in total) whose position on the diagram was discrepant, since they lied far away from the sequence of the cluster, that is clearly outlined. Finally, we also used the $Gaia$ RV to improve our member list. As a result, five other objects have been identified as outliers and, therefore, their membership has been ruled out (see Sect.~\ref{sect_rv}).
Our final member list is composed of 260 objects, out of which 195 are considered type I, 43 belong 
to type II and the remaining 22 are classified as type III members. The cluster probabilities for all our targets are listed in Table~\ref{tab_astrom} and 
the colour-magnitude diagram is displayed in Fig.~\ref{fig_mp2}.

\begin{figure}[!h]
\begin{center}

\includegraphics[width=\columnwidth]{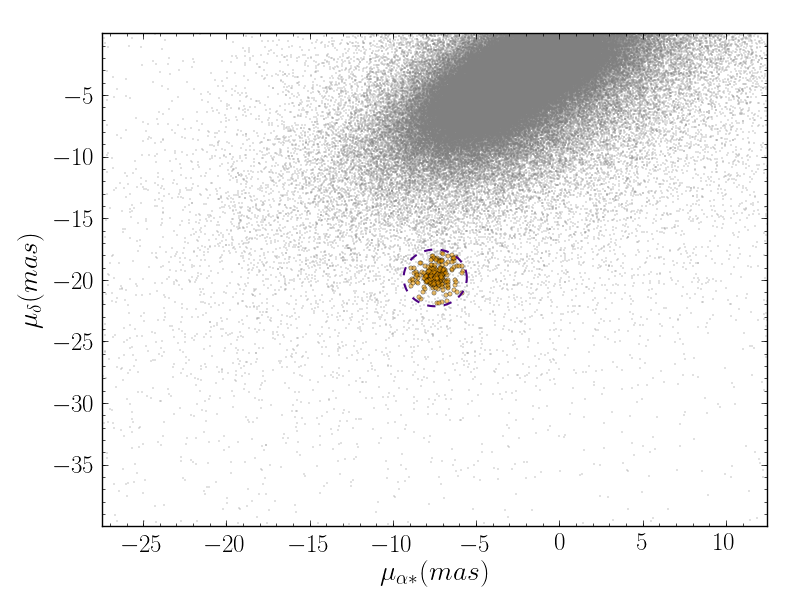}	
\caption{Proper motion diagram in the field of M\,39. The extent of the cluster in the astrometric space is delimited by the ellipse, whose semiaxes correspond
to five times the uncertainty of the mean $\mu_{\alpha*}$, $\mu_{\delta}$ values. Grey dots represent sources with $G\leq$\,16\,mag within a radius of 250' around 
the nominal cluster center.}

\label{fig_mp}
\end{center}
\end{figure}

\begin{figure}[!h]
\begin{center}

\includegraphics[width=\columnwidth]{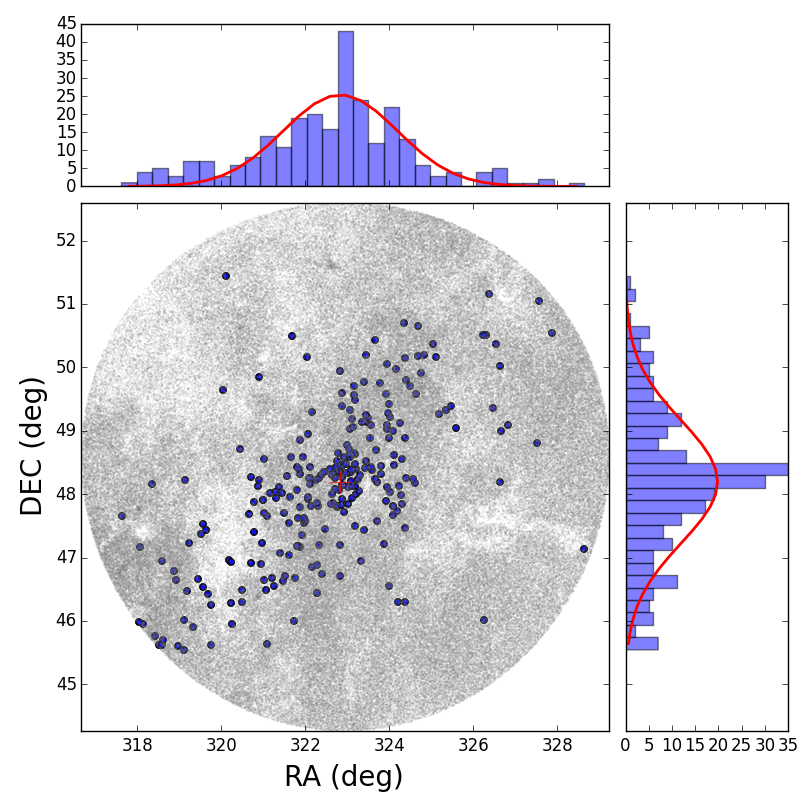}	
\caption{Sky region around M\,39. The density map represents sources with $G\leq$\,16\,mag within a radius of 250' around the nominal cluster centre while the blue circles are the members identified in this work. A histogram showing the member distribution along each axis is also displayed.}

\label{fig_hist}
\end{center}
\end{figure}

We point out that the cluster is well defined in the astrometric space and contamination from field stars is low. We have expanded the member list provided 
by \citet{CG18} by about 30\%, but nevertheless we have not found any giants. According to our criterion we observed 17 high-probability members among our targets, 
while we had to reject the stars 3004, 3423 and 4294 as cluster members. Therefore, their parameters 
were not used in calculating the average values of the cluster.

By examining the spatial distribution of the members in the RA$-$DEC plane (Fig.\ref{fig_hist}) we found the cluster centre by fitting a Gaussian to the histogram showing the projected distribution on each coordinate. In this way we put the cluster at $\alpha$(J2000)\,=\,21$^h$\,31$^m$\,23.4$^s$, 
$\delta$(J2000)\,=\,48$^{\circ}$\,11$\arcmin$\,35.3$\arcsec$, which is slightly displaced with respect to the estimate by \citet{CG18}, ours minus theirs,
$\Delta(\alpha,\delta)$\,=\,($-$10.0$^s$,$-$3.2$\arcmin$).
Likewise, from the parallaxes of the members we derived the distance to the cluster. Firstly, 
we corrected the parallax zero-point offset for each star following the recommendations outlined by \citet{Lindegren21}. As a guide, we found that, on average, 
this correction was of $-$0.032 mas with respect to the published value. We then calculated the individual distances by inverting the corrected parallaxes and finally,
the cluster value was obtained after fitting a Gaussian to the distribution of the member distances.
We place the cluster at 294$\pm$7\,pc, where the error is the $\sigma$ of the Gaussian.

\section{Spectral Analysis}\label{Sec:sp}       

Before starting the spectral analysis, the telluric H$_2$O lines at the H$\alpha$ and Na\,{\sc i}\,D$_2$ wavelengths, as well as those of 
O$_2$ at 6300\,\AA{}, were removed from the spectra. The same procedure described by \citet{Frasca2000} was followed. For this purpose we
used telluric templates (spectra of hot, fast-rotating stars) acquired during the observing run. 

\subsection{Radial velocity}\label{sect_rv}       
    
The first step of our analysis consists in measuring 
the heliocentric radial velocity (RV) of the observed objects. With this aim we cross-correlated the target spectra against synthetic templates by 
using the task {\sc fxcor} of the IRAF\footnote{IRAF is distributed by the National Optical Astronomy Observatory, which is operated by the Association 
of the Universities for Research in Astronomy, Inc. (AURA), under cooperative agreement with the National Science Foundation.} package. 

\begin{table}
\caption{Radial velocities of the newly discovered  SB2 systems.}
\label{Tab:SB2_RV}
\begin{center}
\begin{tabular}{lccc}   
\hline\hline
\noalign{\smallskip}
\multirow{2}{*}{Star}   & HJD  & RV$_1$         & RV$_2$          \\
      & (2\,400\,000+)     & (km\,s$^{-1}$) &  (km\,s$^{-1}$) \\  
\hline   
\noalign{\smallskip}
3423  & 58029.9742   &  $-7.98\pm 1.47$ & $-2.09\pm 0.69$ \\   
3781  & 58055.0448   & $-24.95\pm 0.55$ & $-3.05\pm 4.23$ \\   
3814  & 58055.0556   &  $9.02\pm 1.74$  & $-76.50\pm 9.63$ \\   
\hline   
\end{tabular}
\end{center}
\end{table}

When examining the cross-correlation function (CCF), we note that stars 3423, 3781, and 3814 exhibit a double  or a strongly asymmetric peak, which clearly suggests a binary nature (see Fig.\,\ref{Fig:CCF_Plat3423}). 
However, since we have a single spectrum for each of them in which the peaks are not fully resolved, we cannot provide orbital and stellar parameters but we can only classify them as likely double-lined spectroscopic binaries (SB2s) and provide the radial velocity of their primary (more luminous) and secondary components in Table\,\ref{Tab:SB2_RV}.
For the remaining stars, considered single for all purposes, the RVs are displayed in the last column 
of Table~\ref{tab_params}. 

With the exception of the SB2 systems
and 5609 (a fast rotator), most of the stars have an RV that ranges from $-$6 to $-$3 km\,s$^{-1}$. We obtained 
a mean radial velocity for the cluster, RV=$-$5.46$\pm$0.54. This weighted average has been calculated using as weight $w=1/\sigma_{\rm RV}^2$, where $\sigma_{\rm RV}$ is the RV error of each star. 
The weighted standard deviation has been used to estimate the uncertainty.
We compared our result with the $Gaia$ RVs of our likely members, which were first identified based on astrometry and photometric criteria (Sect.\,\ref{Sec:memb}). Among them, RV is provided for only 122 stars. We derived the cluster mean by fitting a Gaussian to the distribution of the individual values, obtaining RV=$-$6.1$\pm$2.3\,km\,s$^{-1}$. We notice that our result shows strong agreement with $Gaia$ RVs. Additionally, based on these RVs, we rejected five objects as members since their values, within the errors, are beyond 5$\sigma$ the cluster average (see Fig.~\ref{hist_RV}). However, we cannot exclude that these stars are indeed binary systems and the discrepant RV, compared with the cluster mean, is the result of the orbital motion.

\subsection{Atmospheric parameters}\label{sec_params}       
       
The stellar atmospheric parameters of our targets, i.e. effective temperature ($T_{\textrm{eff}}$), surface gravity (log\,$g$) and metallicity ([Fe/H]) have been 
determined by using the {\sf ROTFIT} code \citep{Frasca06}, tailored to our data, as already done in previous works \citep[see e.g., ][]{ASCC123,Stock2}. 
Additionally, {\sf ROTFIT} also provides the projected rotational velocity ($v \sin\,i$) and the spectral type (SpT) for each star. {\sf ROTFIT} works by 
performing a $\chi^2$ minimisation of the difference between the target spectrum and a grid of templates. This difference is evaluated in 28 spectral segments
of 100\,\AA{} each from 4000\,\AA{} to 6800\,\AA{}. Then, the final parameters are calculated by averaging the results of the individual regions, weighting 
them according to the $\chi^2$ and the amount of spectral information contained in each segment. The grid of templates consists of ELODIE spectra ($R$\,=\,42\,000)
of real stars with well-known parameters. It is the same grid used by the Catania node within the $Gaia$-ESO Survey \citep{Smiljanic14,Frasca15}.
Since {\sf ROTFIT} is optimised to analyse FGK-type stars, for the hottest targets we employed a different approach for which we used a grid of synthetic spectra 
computed as described in Sect.~\ref{sec_abund}. For these stars we used the wings and cores of Balmer lines to determine $T_{\textrm{eff}}$ and log\,$g$. The
$v \sin\,i$ was derived in a region around the \ion{Mg}{II}\,$\lambda$4481 line, while a solar metallicity was assumed as the rapid rotation of these sources makes
very difficult to measure it. For more detailed information on our method, the reader is referred to \citet{ASCC123}. Results are summarized in Table~\ref{tab_params}.
The metallicities reported in Table~\ref{tab_params} have been refined by applying {\sf SYNTHE} to our spectra (see Sect.~\ref{sec_abund}) and adopting the [Fe/H] values derived with {\sf ROTFIT} as a starting point. However, these values practically do not change within the errorbars.

\begin{table*}
\caption{Stellar parameters derived for the stars observed in this work.}
\label{tab_params}
\begin{center}
\begin{tabular}{lcccccc}   
\hline\hline
Star    & $T_{\textrm{eff}}$ (K) & $\log\,g$ & [Fe/H] & Sp T  &  $v \sin\,i$ (km\,s$^{-1}$) & RV (km\,s$^{-1}$) \\  
\hline   
\multicolumn{7}{c}{Members}\\    
\hline    
0305  &  5783$\pm$73   &  4.40$\pm$0.13  &     0.03$\pm$0.13  &  G2\,V   &   3.8$\pm$1.2  &   $-$5.34$\pm$0.06  \\   
2451  & 10000$\pm$400  &  4.00$\pm$0.20  &     0.00$^*$          &  A0\,V   & 110.0$\pm$20.0 &   $-$3.15$\pm$0.70  \\
3061  &  6314$\pm$147  &  3.88$\pm$0.12  &  $-$0.40$\pm$0.10  &  F8\,V   &  37.7$\pm$1.7  &   $-$6.07$\pm$0.69  \\
3288  &  5772$\pm$65   &  4.29$\pm$0.10  &     0.13$\pm$0.12  & G5\,IV-V &   4.3$\pm$0.7  &   $-$5.77$\pm$0.07  \\
3311  &  7400$\pm$160  &  4.20$\pm$0.20  &     0.06$\pm$0.11           &  A9\,V   &  42.5$\pm$1.3  &   $-$6.14$\pm$0.23  \\    
4265  & 10100$\pm$400  &  4.00$\pm$0.20  &     0.00$^*$          &  A0\,V   &  52.1$\pm$2.3  &   $-$5.28$\pm$0.64  \\    
4322  &  5782$\pm$69   &  4.44$\pm$0.11  &     0.03$\pm$0.12  &  G2\,V   &   7.5$\pm$0.7  &   $-$5.87$\pm$0.07  \\
4438  &  6233$\pm$138  &  3.89$\pm$0.12  &     0.07$\pm$0.17  &  F7\,IV  &  26.4$\pm$1.5  &   $-$6.63$\pm$0.47  \\
4673  &  8700$\pm$500  &  4.00$\pm$0.20  &     0.00$^*$           &  A3\,V   & 128.4$\pm$8.0  &   $-$3.16$\pm$0.50  \\
5045  &  9100$\pm$200  &  3.50$\pm$0.20  &     0.00$^*$           &  A1\,V   &  51.4$\pm$2.3  &   $-$3.31$\pm$0.46  \\
5609  &  7150$\pm$150  &  4.30$\pm$0.20  &  $-$0.12$\pm$0.12           &  F0\,V   & 142.4$\pm$11.8 &  $-$10.94$\pm$2.02  \\  
5729  &  8000$\pm$400  &  4.00$\pm$0.20  &     0.00$^*$           &  A6\,V   & 190.0$\pm$20.0 &   $-$5.99$\pm$0.49  \\
6791  &  8000$\pm$400  &  4.00$\pm$0.20  &     0.00$^*$           &  A6\,V   & 122.6$\pm$8.2  &   $-$5.68$\pm$0.60  \\
7140  &  9900$\pm$400  &  4.00$\pm$0.20  &     0.00$^*$           &  A0\,V   &  75.0$\pm$10.0 &   $-$4.10$\pm$0.75  \\
ucac  &  5783$\pm$78   &  4.41$\pm$0.12  &     0.06$\pm$0.10  &  G1\,V   &   6.1$\pm$1.2  &   $-$4.78$\pm$0.08  \\
\hline   
\multicolumn{7}{c}{Non-members}\\    
\hline 
3004  &  6921$\pm$270  &  4.22$\pm$0.15  &     0.08$\pm$0.14  &  F2\,V   &  25.7$\pm$2.5  &   $-$5.73$\pm$0.48  \\
4294  & 10200$\pm$350  &  4.00$\pm$0.20  &     0.00$^*$           & B9.5\,V  & 110.0$\pm$20.0 &   $-$2.10$\pm$0.80  \\
\hline
 
\end{tabular}
\end{center}
\begin{list}{}{}
\item[$^*$] Solar metallicity has been adopted for the hottest stars of our sample, as explained in the text.
\end{list}
\end{table*}

\subsection{Chemical abundances}\label{sec_abund}       

We calculated the elemental chemical abundances of our targets following the same procedure adopted in some of our previous works related to the SPA project
\citep[see, e.g., ][]{ASCC123,Stock2}, which is
based on the spectral synthesis technique \citep{Catanzaro11,Catanzaro13}. In a first step we computed 1D local thermodynamic equilibrium (LTE) atmospheric models
with the {\sf ATLAS9} code \citep{Kurucz1993a,Kurucz1993b}, adopting the {\sf ROTFIT} atmospheric parameters (Sect.~\ref{sec_params}). Then, we generated 
the corresponding synthetic spectra using {\sf SYNTHE} as the radiative transfer code \citep{Kurucz1981}. We compared the observed spectra with the synthetic
ones (adequately broadened taking into account both the instrumental profile and the rotational one) in 39 spectral segments of 50\,\AA{} each between 4400 and 6800\,\AA{}.
Finally, by minimising the $\chi^2$ of their differences, the chemical abundances are found. In this way, for those stars with $T_{\textrm{eff}}$<7500\,K, nine 
in total, we investigated 21 elements with atomic number up to 56 namely, C, Na, Mg, Al, Si, S, Ca, Sc, Ti, V, Cr, Mn, Fe, Co, Ni, Cu, Zn, Sr, Y, Zr, and Ba.
Individual abundances are listed according to the standard notation $A(X)$=log[$n(X)/n(H)$]+12 in Table~\ref{tab_ind_abund} and an example of the spectral synthesis is shown in Fig.~\ref{spectra}. The cluster average composition, relative 
to solar abundances by \citet{Grevesse07}, is also reported in Table~\ref{tab_abund_avrg}. For each element, it has been calculated by taking the the weighted average of the values for each star, using the individual errors as weights. The uncertainties express, in terms of standard deviation, the dispersion of stellar abundances around the cluster value.

\begin{figure}[ht]
\begin{center}

\includegraphics[width=\columnwidth]{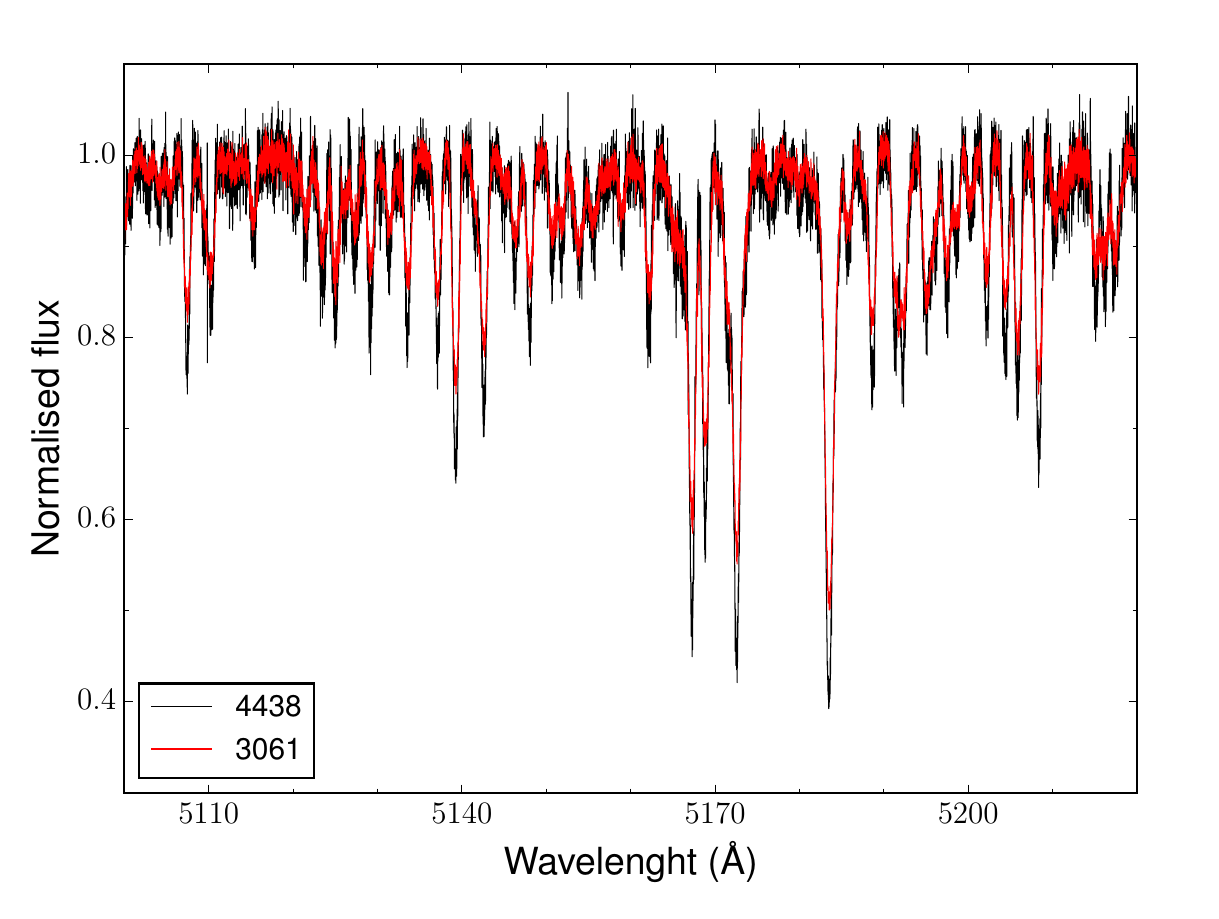}	
\caption{Comparison of the spectra of two cluster stars with similar stellar parameters in the region of \ion{Mg}{i}b triplet. Star 4438 shows a chemical composition similar to that of other cluster members while 
star 3061 exhibits a somewhat lower value, which is also suggested by its shallower lines.} 

\label{fig_comp_metalic}
\end{center}
\end{figure}

We find that the cluster, on average, exhibits a solar-like composition, with a metallicity of [Fe/H]=$+$0.04$\pm$0.08\,dex. Only Na shows a significant subsolar 
abundance ([Na/Fe]=$-$0.25$\pm0.13$). For the $\alpha$-elements the weighted-mean ratio is [$\alpha$/Fe]=$+$0.12$\pm$0.10\,dex, with S being overabundant with respect to the other elements ([S/Fe]=$+$0.30$\pm$0.12).
The Fe-peak elements show a solar ratio, [X/Fe]=$+$0.05$\pm$0.10\,dex, while the neutron-capture ones, especially Ba, have a supersolar value, [$s$/Fe]=$+$0.28$\pm$0.10\,dex. 
As expected, a homogeneous chemical composition is observed among the cluster members, with the only exception of star 3061, which in general displays lower
abundances (i.e.  Fe and Ti, see Fig.~\ref{fig_comp_metalic}). On the contrary, star 3004, which was discarded as a member (see Sect.~\ref{Sec:memb}), has a chemical
composition compatible with that of the cluster.

\begin{table}
\caption{Average chemical composition of M\,39, relative to solar abundances by \citet{Grevesse07}, derived in this work.}
\centering
\begin{tabular}{cc|cc}
\hline
\hline
X   &     [X/H]            &  X  &         [X/H]       \\ 
\hline
C   &  $+$0.10 $\pm$ 0.15  &  Mn &  $+$0.10 $\pm$ 0.10  \\
Na  &  $-$0.21 $\pm$ 0.11  &  Fe &  $+$0.04 $\pm$ 0.08  \\
Mg  &  $+$0.12 $\pm$ 0.07  &  Co &  $+$0.05 $\pm$ 0.06  \\
Al  &  $+$0.18 $\pm$ 0.06  &  Ni &  $+$0.00 $\pm$ 0.05  \\ 
Si  &  $+$0.07 $\pm$ 0.09  &  Cu &  $+$0.15 $\pm$ 0.07  \\
S   &  $+$0.34 $\pm$ 0.09  &  Zn &  $-$0.03 $\pm$ 0.19  \\
Ca  &  $+$0.19 $\pm$ 0.07  &  Sr &  $+$0.32 $\pm$ 0.08  \\
Sc  &  $+$0.11 $\pm$ 0.07  &  Y  &  $+$0.27 $\pm$ 0.02  \\ 
Ti  &  $+$0.11 $\pm$ 0.15  &  Zr &  $+$0.29 $\pm$ 0.08  \\ 
V   &  $+$0.22 $\pm$ 0.05  &  Ba &  $+$0.41 $\pm$ 0.04  \\ 
Cr  &  $+$0.15 $\pm$ 0.04  &     &                      \\

\hline
\end{tabular}
\label{tab_abund_avrg}
\end{table}

\section{SED fitting and reddening}\label{Sec:sed}

\begin{figure}[ht]
\begin{center}

\includegraphics[width=\columnwidth]{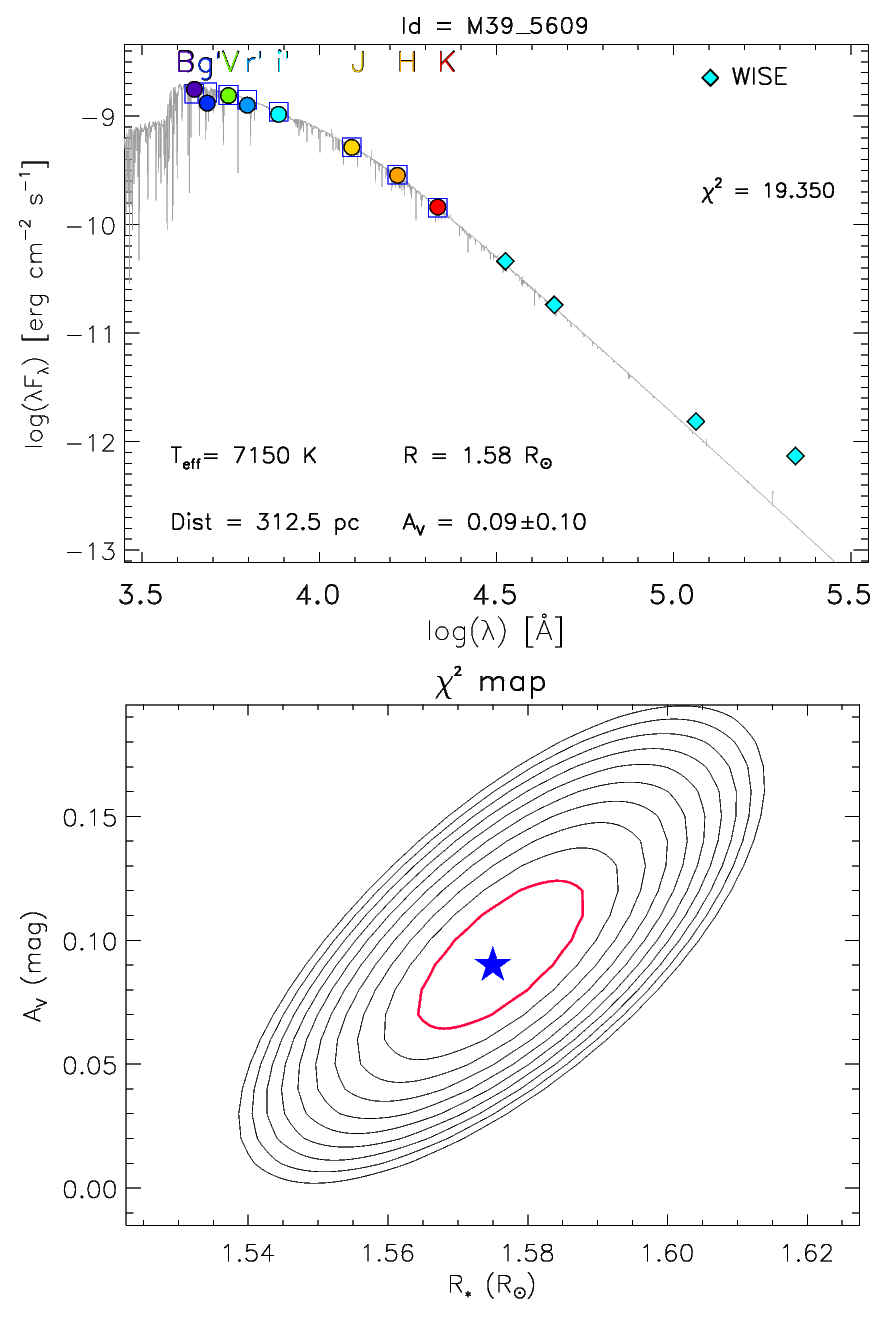}	
\caption{\textit{Top:} SED fitting of the star 5609. \textit{Botton:} $\chi^2$-contour map of the fitting. The red line corresponds to the 1$\sigma$ confidence
level}

\label{fig_sed}
\end{center}
\end{figure}

In order to evaluate how our observations are affected from interstellar extinction ($A_V$), we resorted to the spectral energy distribution (SED) fitting tool, as
previously done in other works \citep{ASCC123,Stock2}. For each of our targets we fitted the corresponding SED with BT-Settl synthetic spectra \citep{Allard14}.
To build the SED, on the one hand we made use of publicy available optical and near-infrared photometric data. Specifically, we used the $BVg'r'i'$
\citep[from the APASS catalogue,][]{APASS}, $JHK_{\textrm{S}}$ \citep[2MASS,][]{2MASS} and $W_1W_2W_3W_4$ bands \citep[WISE, ][]{WISE}. 
On the other hand, we adopted the atmospheric parameters ($T_{\textrm{eff}}$ and log\,$g$) derived in Sect.~\ref{sec_params} as well as the (corrected) $Gaia$-DR3 parallax. The stellar radius ($R$) and the $A_V$ were set as free parameters. Their best values were obtained by $\chi^2$ minimisation. The errors on 
$A_V$ and $R$ are found by the minimisation procedure considering the 1$\sigma$ confidence level of the $\chi^2$ map, but we also took the error on the
$T_{\textrm{eff}}$ into account. An example of this fitting is shown in Fig.~\ref{fig_sed}. 

The results of the SED fitting are reported in Table~\ref{tab_sed}. Among the members analysed the $A_V$ ranges from 0.00 to 0.38 mag, with an average of 
$A_V$=0.06$\pm$0.10, where the error is the standard deviation. For most 
stars the reddening is on the cluster average with only star 7140 showing a 
slightly different value, 0.38\,mag, the largest in the sample. If we recalculate the average without taking this star into account we obtain $A_V$=0.04$\pm$0.05\,mag,
where the dispersion has decreased. 
This result indicates that extinction is negligible in
the cluster field, in good agreement with
the literature \citep[$A_V=0.0$\,mag]{CG18}.

\begin{table}
\caption{Results of the SED fitting.}
\label{tab_sed}
\begin{center}
\begin{tabular}{lccc}  
\hline\hline
Star    & $A_V$ (mag) & $R$ (R$_{\odot}$) & $L$ (L$_{\odot}$) \\  
\hline  
\multicolumn{4}{c}{Members}\\    
\hline    
0305  &  0.00$\pm$0.05  &  0.92$\pm$0.01  &   0.8$\pm$0.1  \\
2451  &  0.11$\pm$0.07  &  3.56$\pm$0.07  & 113.9$\pm$18.2 \\
3061  &  0.00$\pm$0.03  &  1.24$\pm$0.01  &   2.2$\pm$0.2  \\
3288  &  0.04$\pm$0.07  &  0.80$\pm$0.01  &   0.6$\pm$0.1  \\
3311  &  0.00$\pm$0.08  &  1.47$\pm$0.01  &   5.9$\pm$0.5  \\
4265  &  0.09$\pm$0.07  &  2.49$\pm$0.04  &  55.6$\pm$8.8  \\
4322  &  0.03$\pm$0.07  &  0.93$\pm$0.01  &   0.9$\pm$0.1  \\
4438  &  0.00$\pm$0.01  &  1.24$\pm$0.07  &   2.1$\pm$0.3  \\
4673  &  0.00$\pm$0.03  &  1.89$\pm$0.11  &  18.4$\pm$4.7  \\
5045  &  0.16$\pm$0.12  &  5.10$\pm$0.13  & 160.8$\pm$14.4 \\
5609  &  0.09$\pm$0.10  &  1.58$\pm$0.02  &   6.0$\pm$0.5  \\
5729  &  0.00$\pm$0.01  &  1.87$\pm$0.13  &  13.0$\pm$3.1  \\
6791  &  0.01$\pm$0.16  &  1.64$\pm$0.04  &   9.9$\pm$2.0  \\
7140  &  0.38$\pm$0.11  &  1.62$\pm$0.04  &  22.6$\pm$3.6  \\
ucac  &  0.06$\pm$0.10  &  0.87$\pm$0.01  &   0.8$\pm$0.1  \\
\hline  
\multicolumn{4}{c}{Non-members}\\    
\hline  
3004  &  0.00$\pm$0.03  &  1.49$\pm$0.03  &   4.6$\pm$0.7  \\
4294  &  0.55$\pm$0.13  &  6.10$\pm$0.06  & 335.0$\pm$46.7 \\
\hline
 
\end{tabular}
\end{center}
\end{table}  

\section{Colour-magnitude diagrams}\label{Sec:CMD}

To investigate the cluster age, we resorted to the 
 isochrone-fitting method. In a first step we used archival photometry 
to plot the colour-magnitude diagram (CMD) of the cluster, which is a snapshot of its evolutionary stage. Then, we looked for the age-dependent model, the isochrone, that best 
reproduced the position of the cluster members on the CMD. To compute the different isochrones we took advantage of the results previously obtained from spectroscopy,
that is metallicity (Sect.~\ref{sec_abund}) and reddening (Sect.~\ref{Sec:sed}).

We used three different photometric systems (Johnson, 2MASS and $Gaia$) to construct the following diagrams: $M_V$/$(B-V)_0$, 
$M_{K_{\textrm{S}}}$/$(J-K_{\textrm{S}})_0$ and $M_G$/$(G_{\textrm{BP}}-G_{\textrm{RP}})_0$. In each of the diagrams the number of plotted members  
is different (111, 
252 and 265, respectively) since photometric data is not always available for all of them. To convert apparent to absolute magnitudes we used the individual 
distance of each source. We calculated it as the inversion of the parallax, once it was corrected as explained above (Sect.~\ref{Sec:memb}). 

\begin{figure*}[ht]
\begin{center}

\includegraphics[width=17cm]{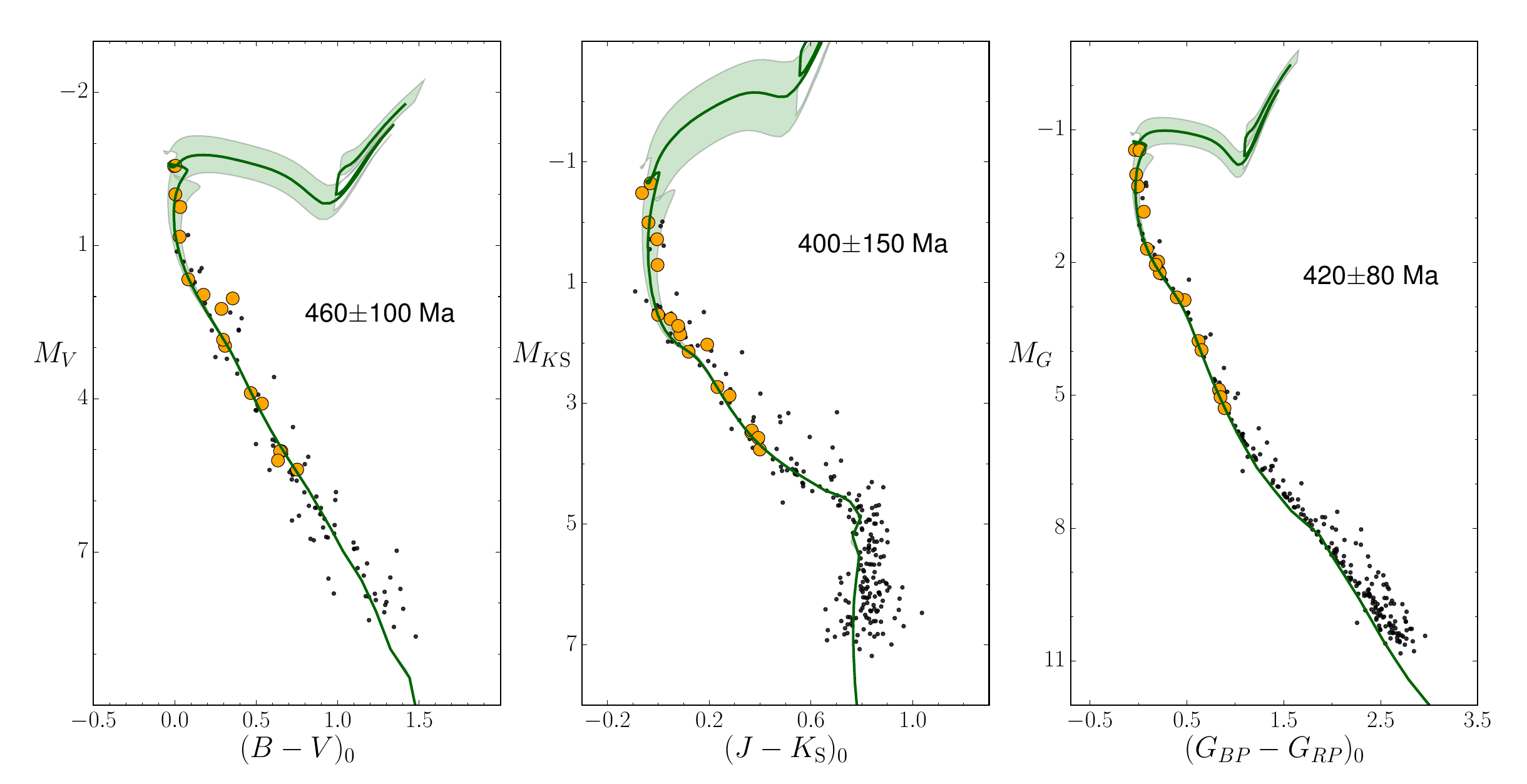}	
\caption{Colour-magnitude diagrams for M39 in three different photometric systems. \textit{Left:} $M_V$/$(B-V)_0$, photometric data from the APASS catalogue. 
\textit{Centre:} $M_{K_{\textrm{S}}}$/$(J-K_{\textrm{S}})_0$ (2MASS). \textit{Right:} $M_G$/$(G_{\textrm{BP}}-G_{\textrm{RP}})_0$ ($Gaia$-DR3). 
The small black points are the members selected in this work and the orange circles mark those ones observed spectroscopically. The green lines
and the shaded area are the best-fitting isochrones with the associated uncertainties. The age derived from each CMD is also shown. 
}

\label{fig_CMD}
\end{center}
\end{figure*}

In the two first CMDs the reddening was corrected individually for those stars observed in this work with the value previously derived, while for the
remaining stars the mean cluster value was applied. In the $Gaia$ CMD we made use of the $A_G$ and $E(G_{\textrm{BP}}-G_{\textrm{RP}})$ published in the DR2.
When comparing all the three CMDs we immediately realised that the brightest stars in the upper MS behaved anomalously in the $M_V$/$(B-V)_0$ diagram compared 
to the other two. This is due to the fact that these stars (with $V\approx$7\,mag) are likely saturated in the APASS photometry. In order to fix this issue we 
resorted to the ASCC2.5 catalogue \citep{Kharchenko09}, from which we took $V$ and ($B-V$) for six stars brighter than $V$=8 after scaling both photometric 
datasets\footnote{We selected 25 members with good photometry in both catalogues in the range 7.5$\leq$\,$V$\,$\leq$10.5, finding average differences (ASCC2.5 
minus APASS) of $\Delta\,V$=0.009 and $\Delta\,(B-V)$=$-$0.019.}. The distribution of these stars on the CMD improved considerably after this correction, so that we decided to use these magnitudes (listed in Table\,\ref{tab_fotom} and displayed in Fig.\,\ref{fig_CMD}) in this work. 

Once the dereddened CMDs were built, we overplotted on them a set of isochrones of different ages to find the one that best reproduced the distribution of the
cluster members and, thus, find the likely age for the cluster. We used PARSEC isochrones \citep{Marigo17} computed at the metallicity found in this work
([Fe/H]=$0.04$) in a wide range of ages (log\,$\tau$=8.2--8.9, which corresponds to ages from 160 to 780 Ma). We carefully examined by eye how well the isochrones 
(in steps $\Delta$\,log$\tau$=0.01) fitted the CMDs. Since there are no giants we had to pay attention to the stars located at the upper part of the MS. We noted 
that the best-fitting isochrone is not exactly the same in all the diagrams, but perfectly compatible within the errors.
From the optical CMD we find an age for the cluster of 460$\pm$100\,Ma, similar to that obtained from the infrared one, 400$\pm$150\,Ma. The errors represent the interval of isochrones that are capable of reproducing the CMD.

In the case of the $Gaia$ CMD, it is worth highlighting the role that reddening plays. As previously commented, we started taking the $Gaia$ data for the reddening and 
extinction. When available we used 
individual values, as we did in previous works. 
However, for the rest of members we estimated these two values by averaging those of 
the five nearest stars (using their distances as weights). Nevertheless, the average reddening that can be deduced for our members from the $Gaia$ DR3 catalogue, namely
$A_G$=0.30$\pm$0.23 and $E(G_{\textrm{BP}}-G_{\textrm{RP}})$=0.20$\pm$0.14, is quite different from the one calculated by us and exhibits a greater dispersion. When adopting it, the age of 
cluster becomes $\approx$300\,Ma, somewhat younger than the age derived from the other CMDs. If we instead redden the isochrone with the $A_V$ derived in this 
work, the dispersion of the stars along the MS decreases and the isochrones fit better the CMD. In this way, the cluster turns out to be of 420$\pm$80, a value 
comprised between those derived from the other two CMDs. This difference between our extinction and that derived from $Gaia$ data could be explained by the known fact that $Gaia$-DR3 tends to overestimate the extinction in areas of the sky where its value is very low \citep{Andrae23, Babusiaux23, Fouesneau23}.

Finally, the definitive age of the cluster is obtained by averaging the estimates derived from each CMD, using the individual uncertainties as weights. The resulting 
value is 430$\pm$110\,Ma, which corresponds to a MSTO mass of around 2.8\,M$_{\sun}$. As error we took the mean of each uncertainty. The CMDs, along with the best-fitting 
isochrones and their uncertainties, are shown in Fig.~\ref{fig_CMD}. This result agrees very well
 with that reported in \citet{CG20}.

\section{Discussion}\label{Sec:disc}

\subsection{Radial and rotational velocities}

M\,39 was one of the nearby clusters containing FK dwarfs observed by \citet{Mermilliod09}. They derived radial and rotational velocities for 24 stars in the 
cluster field, out of which six stars are also included in our sample. Star 3423 is one of the targets in common in both studies, but in our spectrum it appears as an SB2, with two strongly blended peaks in the CCF, whose velocities are RV$_1=-7.98\pm 1.47$\,\kms\ and RV$_2=-2.09\pm 0.19$\,\kms, with the former having a much larger width  
compared to the latter (see Fig.~\ref{Fig:CCF_Plat3423}). 
\citet{Mermilliod09} did report no binary signature for this star. This has likely arisen from having observed the system near to a conjunction. In fact, they measured an RV=$-$5.90$\pm$0.73\,km\,s$^{-1}$, that, if considered as the RV of centre of mass of the system, is in good agreement with the mean of the RVs of the two components in our spectrum. 
For the remaining five objects 
both sets of measurements, RV and  $v \sin\,i$, are in excellent agreement in both papers (see Table~\ref{tab_comp_Me09}). In addition, the mean RV of the cluster
calculated by \citet{Mermilliod09} from 17 members, RV=$-$5.26$\pm$0.30\,km\,s$^{-1}$, is fully compatible with that derived in this work.

\begin{table}
\caption{Radial and rotational velocities for the stars observed by \citet{Mermilliod09} in common with this work (TW).}
\centering
\begin{tabular}{c|cc|cc}
\hline
\hline
\multirow{2}{*}{Star}  &  \multicolumn{2}{c|}{Me09}    & \multicolumn{2}{c}{TW}    \\
                       &     RV            &  $v \sin\,i$  &  RV    &   $v \sin\,i$  \\
\hline
0305  &  $-$5.01$\pm$0.27  &   6.3$\pm$2.8  &  $-$5.34$\pm$0.06  &   4.0$\pm$0.9  \\
3004  &  $-$6.91$\pm$0.76  &  23.7$\pm$2.5  &  $-$5.73$\pm$0.48  &  25.7$\pm$2.5  \\
3061  &  $-$5.14$\pm$0.90  &  40.0$\pm$4.0  &  $-$6.07$\pm$0.69  &  37.7$\pm$1.7  \\
3288  &  $-$5.76$\pm$0.40  &   5.8$\pm$2.1  &  $-$5.77$\pm$0.07  &   4.3$\pm$0.7  \\ 
4322  &  $-$6.45$\pm$0.22  &   4.9$\pm$2.0  &  $-$5.87$\pm$0.07  &   7.5$\pm$0.7  \\

\hline
\end{tabular}
\label{tab_comp_Me09}
\end{table}

\subsection{Chromospheric emission and lithium abundance}

A way of estimate the age of the stars that are forming the cluster, independent of the isochrone-fitting method, is the evaluation of the level of magnetic
activity (e.g., the emission in the cores of lines formed in the chromosphere) and the abundance of Li present in their atmospheres. This is valid for stars 
cooler than $\approx$\,6500\,K and ages between a few ten and a few hundred Ma \citep[see][and references therein]{Jeffries14,Frasca18}.

\begin{figure}
\begin{center}
\vspace{0cm}
\includegraphics[width=\columnwidth]{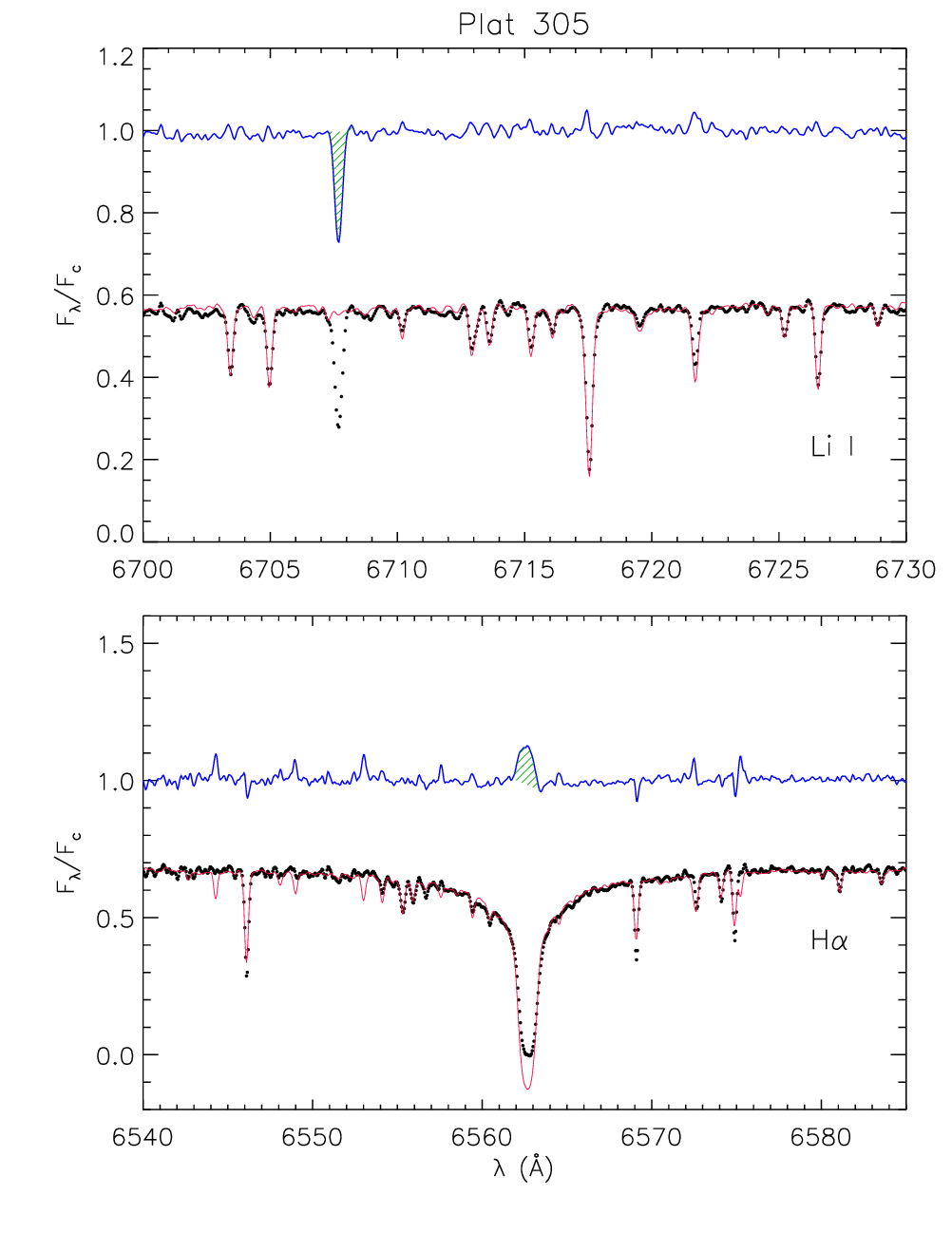}
\vspace{0cm}
\includegraphics[width=\columnwidth,height=4.5cm]{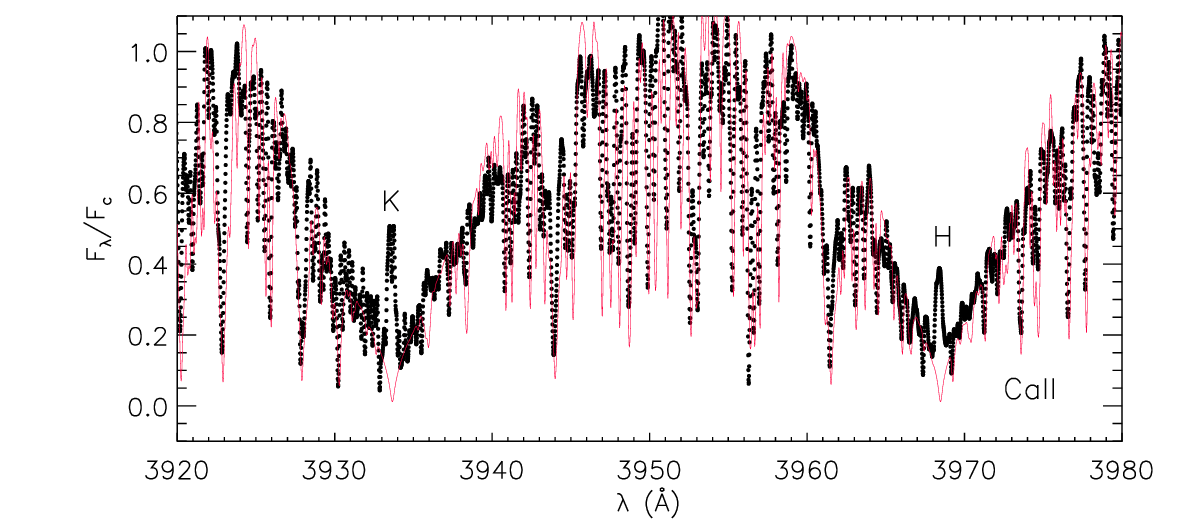}
\caption{Spectral subtraction for the star 0305 in the \ion{Li}{i}\,$\lambda$6708\,\AA, H$\alpha$, and \ion{Ca}{ii}\,H\&K lines ({\it from top to bottom}).}

\label{fig_sub}
\end{center}
\end{figure}

In order to perform this task we applied the subtraction technique. It consists in subtracting the spectrum of a pertinent template from that of the target 
to measure the excess emission in the chromospheric lines and the equivalent width of the \ion{Li}{I}\,$\lambda$\,6708\,$\AA$ absorption line ($EW_{\textrm{Li}}$).
An example of this procedure is shown in Fig.~\ref{fig_sub}.
As templates we used
the spectra of non-active and lithium-poor stars, 
Doppler-shifted and rotationally broadened by {\sf ROTFIT}. As diagnostics 
of activity we used the Balmer H$\alpha$ line and the \ion{Ca}{ii} H$\&$K lines. The latter lines are located in the bluest part of the spectra where
the S/N is very low, but degrading the spectra resolution from 115,000 to that of ELODIE ($R=42,000$) allowed us to improve the S/N and apply the spectral subtraction. For the \ion{Ca}{ii} lines, the photospheric templates are made with synthetic BT-Settl spectra \citep{Allard14} as we did in \citet{Frasca23a}, because they are best suited to reproduce the photospheric spectrum without any chromospheric contribution compared to the low-activity real stars. Indeed, for the latter a small line emission or filling is very often present in the cores of the \ion{Ca}{ii} H\&K lines. 

\begin{figure}[htb]
\begin{center}
\includegraphics[width=8.0cm]{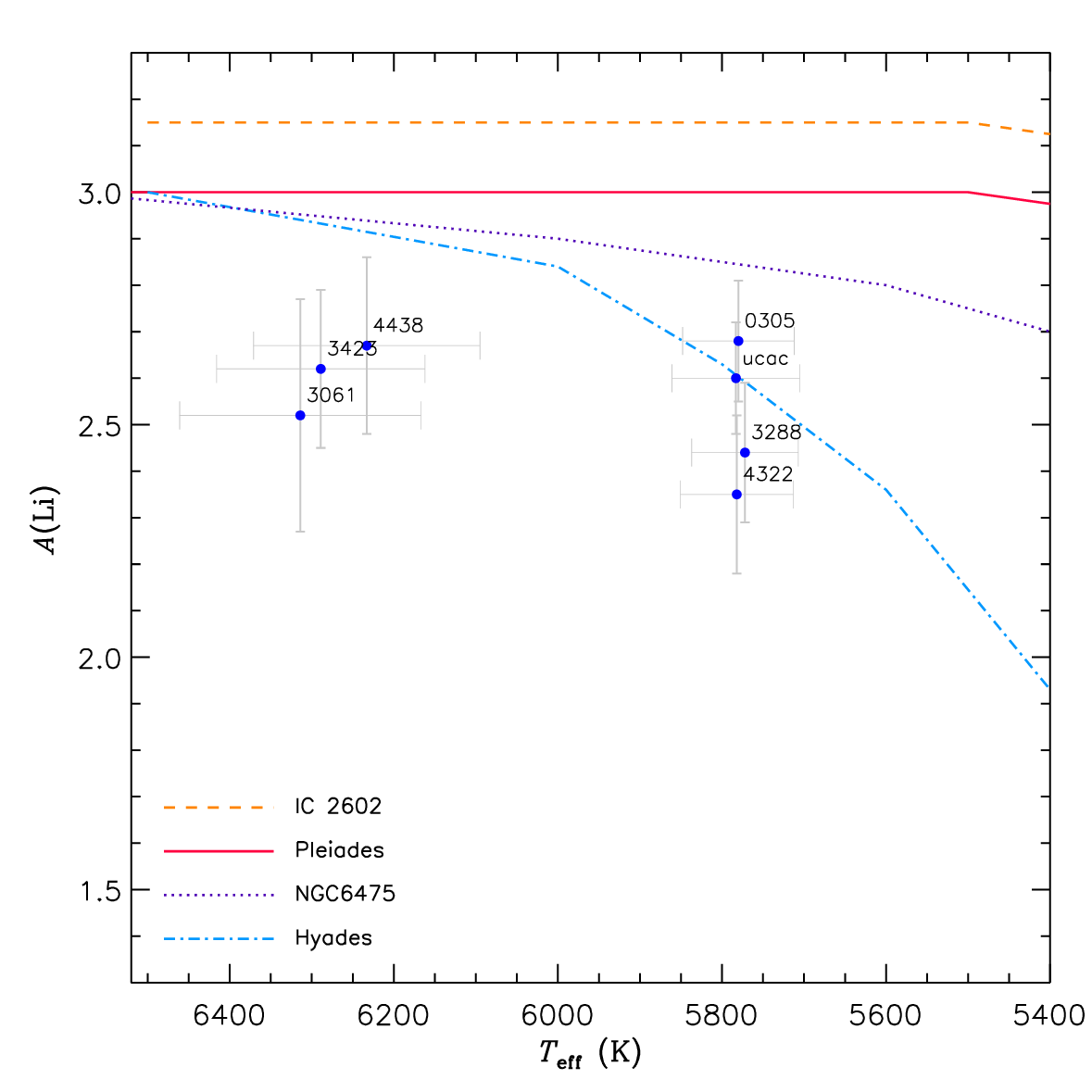}           \caption{Lithium abundance as a function of \teff\ for the solar-type members (\teff$<$6500\,K).
The upper envelopes of $A$(Li) for four young clusters  
adapted from \citet{Sestito2005} are overplotted. 
}
\label{Fig:NLi}
\end{center}
\end{figure}

From the atmospheric parameters and the $EW_{\textrm{Li}}$ we estimated the lithium abundance, $A$(Li), by interpolating the curves of growth of \citet{Soderblom93},
calculated in the $T_{\textrm{eff}}$ range 4000--6500\,K. 
The lithium abundance is reported in Table~\ref{tab_li} and is plotted versus \teff\ in Fig.~\ref{Fig:NLi}. The upper envelopes of four young clusters, namely the Hyades (650 Ma), NGC\,6475 (300 Ma),
the Pleiades (125 Ma), and IC\,2602 (30 Ma), adapted from \citet{Sestito2005}, are overlaid with different colors.
All the solar-type members display values of $A$(Li) compatible with an age greater than 300\,Ma. 

\begin{table}
\caption{\ion{Li}{I}\,$\lambda$\,6708\,$\AA$ equivalent width and abundance for the cool stars observed in this work.}
\begin{center}
\begin{tabular}{lccc}
\hline
\hline
\multirow{2}{*}{Star}  & $T_{\textrm{eff}}$       &  $EW_{\textrm{Li}}$ & \multirow{2}{*}{$A$(Li)}\\
                       &       (K)       &  (m$\AA$)   &        \\ 
\hline
0305   &  5783$\pm$73   &  100$\pm$10  &  2.68$\pm$0.13     \\
3004   &  6921$\pm$270  &   29$\pm$9   &  2.63$\pm$0.14$^*$  \\
3061   &  6314$\pm$147  &   31$\pm$9   &  2.52$\pm$0.25      \\ 
3288   &  5772$\pm$65   &   69$\pm$11  &  2.44$\pm$0.15      \\
3311   &  7088$\pm$160  &   28$\pm$6   &  2.60$\pm$0.11$^*$  \\
4322   &  5782$\pm$69   &   57$\pm$11  &  2.35$\pm$0.17      \\ 
4438   &  6233$\pm$138  &   50$\pm$8   &  2.67$\pm$0.19      \\
5609   &  7029$\pm$167  &   4$\pm$4   &     < 2.03$^*$      \\
ucac   &  5783$\pm$78   &  90$\pm$7   &  2.60$\pm$0.12      \\

\hline
\end{tabular}
\end{center}
\begin{list}{}{}
\item[$^*$] $T_{\textrm{eff}}$\,>\,6500\,K. $A$(Li) extrapolated from the \citet{Soderblom93} tables. 
\end{list}
\label{tab_li}
\end{table}

\begin{figure}
\begin{center}
\includegraphics[width=\columnwidth]{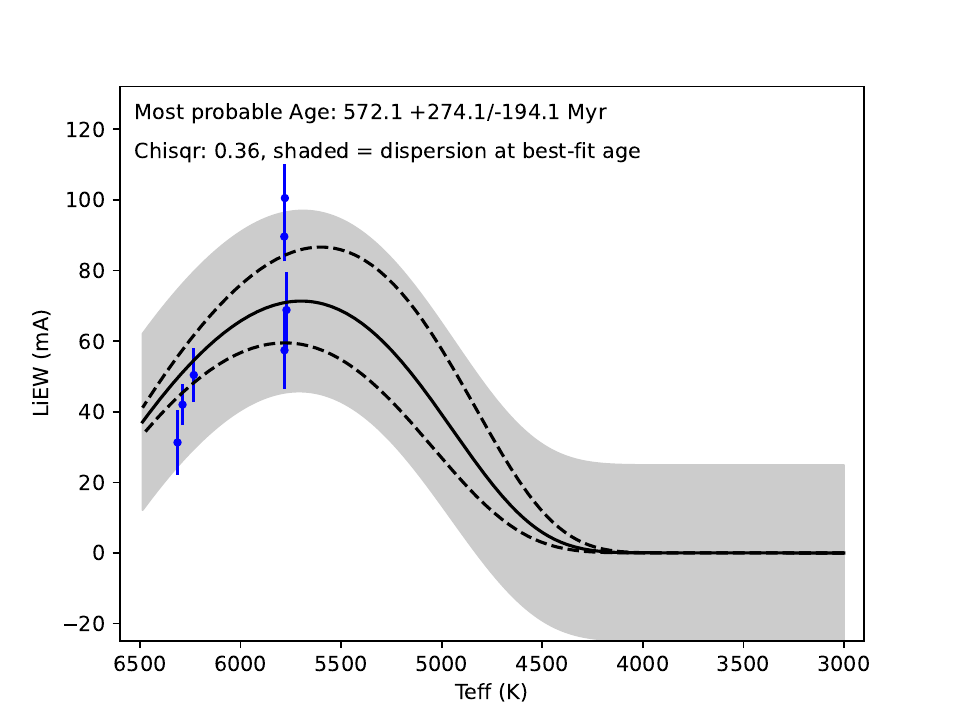}
\caption{Fit to the Li depletion pattern of the cool members of M\,39 made with the {\sf EAGLES} code.}
\label{fig_eagles}
\end{center}
\end{figure}

To estimate the age of the cluster from the content of photospheric lithium, we also uses
the {\sf EAGLES} code \citep{Jeffries23}, as we
did in \citet{Frasca23b}. This code, based on an empirical model, is very suitable for
a cluster since it determines the age probability 
distribution for the ensemble from the $T_{\textrm{eff}}$ and $EW_{\textrm{Li}}$ of its members. In the present case, we obtained an age for M\,39 of 
$\tau$=570$^{+270}_{-190}$\,Ma (see Fig.~\ref{fig_eagles}). This value is greater than the one obtained with the fitting-isochrone method, but compatible within 
the errors. In any case, the age uncertainty is larger than one could expect, mainly due to the data scatter and the poor $T_{\rm eff}$ distribution of the targets, and it does not allow us to provide us with a good age constraint.

The excess H$\alpha$ equivalent width, $EW_{\rm H\alpha}^{em}$, has been obtained by integrating the H$\alpha$ emission profile, resulting by the subtraction of the photospheric template, which
is highlighted by the green hatched area in the difference spectra of Fig.~\ref{fig_sub}.
Similarly, the difference between the observed and template spectrum in the \ion{Ca}{ii} region leaves emission excesses in the cores of the H and K lines, whose integration provide the equivalent widths, $EW_{\rm CaII-H}$ and $EW_{\rm CaII-K}$. 
More effective indicators of chromospheric activity are the chromospheric line fluxes at the stellar surface, $F_{\rm line}$, and the luminosity ratios, $R'_{\rm line}=F_{\rm line}/F_{\rm bol}$, where $F_{\rm bol}=\sigma T_{\rm eff}^4$ is the bolometric flux.
For the \ion{Ca}{ii}\,K, the line flux can be calculated as

\begin{equation}
F_{\rm CaII-K}  =  F_{3933}EW_{\rm CaII-K}, 
\end{equation}

{\noindent where $F_{3933}$ is the flux at the continuum at the center of the \ion{Ca}{ii}-K line per unit stellar surface area, which is evaluated 
from the BT-Settl spectra \citep{Allard14} at the stellar temperature and surface gravity of the target. 
The flux error include the error in the equivalent width
and the uncertainty in the continuum flux at the line center,
which is estimated propagating the errors of $T_{\rm eff}$ and $\log g$. }

We computed surface fluxes for the other chromospheric diagnostics in the same way as for the \ion{Ca}{ii}-K line. The EWs, fluxes, and luminosity ratios are reported in Table~\ref{Tab:Halpha_CaII} for the solar-type stars (\teff$<6500$\,K).

We used the age--activity relation proposed by \citet{Mamajek2008} 
to estimate the age of M\,39. From their Eq.\,3 and our values of $R'_{\rm HK}$, which are reported in Table~\ref{Tab:Halpha_CaII}, we estimate an age of 250$\pm$150\,Ma, which is smaller than the values derived from both the CMD and lithium isochrones, but marginally compatible with them if we consider the large age errors.  

\begin{table*}
\caption{H$\alpha$ and \ion{Ca}{ii} H\,\&\,K equivalent widths and fluxes.}
\begin{center}
\begin{tabular}{lcccccccc}
\hline
\noalign{\smallskip}
\multirow{2}{*}{Star}      & $EW_{\rm H\alpha}$  & $F_{\rm H\alpha}$ & \multirow{2}{*}{$R'_{\rm H\alpha}$} & $EW_{\rm CaII-K}$  & $EW_{\rm CaII-H}$  & $F_{\rm CaII-K}$  & $F_{\rm CaII-H}$ & \multirow{2}{*}{$R'_{\rm HK}$} \\   
      & (m\AA) & ($10^5$erg\,cm$^{-2}$s$^{-1}$) & & (m\AA) & (m\AA)  & ($10^5$erg\,cm$^{-2}$s$^{-1}$) & ($10^5$erg\,cm$^{-2}$s$^{-1}$) & \\  
\hline
\noalign{\smallskip}
0305  &  126$\pm$21  &  9.2$\pm$1.6 & $-$4.84 & 262$\pm$88 & 213$\pm$83 & 15.0$\pm$5.3 & 12.2$\pm$4.9 & $-$4.37\\
3061   &  59$\pm$15  & 5.9$\pm$1.6  & $-$5.18 & 157$\pm$40 & 110$\pm$44  & 18.7$\pm$5.9 & 13.2$\pm$5.8 &  $-$4.45 \\ 
3288   & 111$\pm$22  & 8.2$\pm$1.7  &  $-$4.88 & 410$\pm$226 & 194$\pm$226 & 23.1$\pm$12.9 & 11.0$\pm$12.8 & $-$4.27 \\
4322   & 200$\pm$34  & 14.9$\pm$2.6  & $-$4.63 & 331$\pm$130  & 225$\pm$167 & 18.9$\pm$7.6 & 12.8$\pm$9.6 & $-$4.30\\ 
4438   &  96$\pm$14  & 9.4$\pm$1.6  &  $-$4.96 & 207$\pm$60  & 181$\pm$81  & 22.4$\pm$7.5 & 19.6$\pm$9.4 & $-$4.31\\
ucac   & 146$\pm$22  & 10.7$\pm$1.7  &  $-$4.77 & 256$\pm$101 & 230$\pm$115 & 14.7$\pm$6.0  & 13.1$\pm$6.7 & $-$4.37 \\

\hline
\end{tabular}
\end{center}
\label{Tab:Halpha_CaII}
\end{table*}

\subsection{Galactic metallicity gradient}

\begin{figure}[ht]
\begin{center}

\includegraphics[width=\columnwidth]{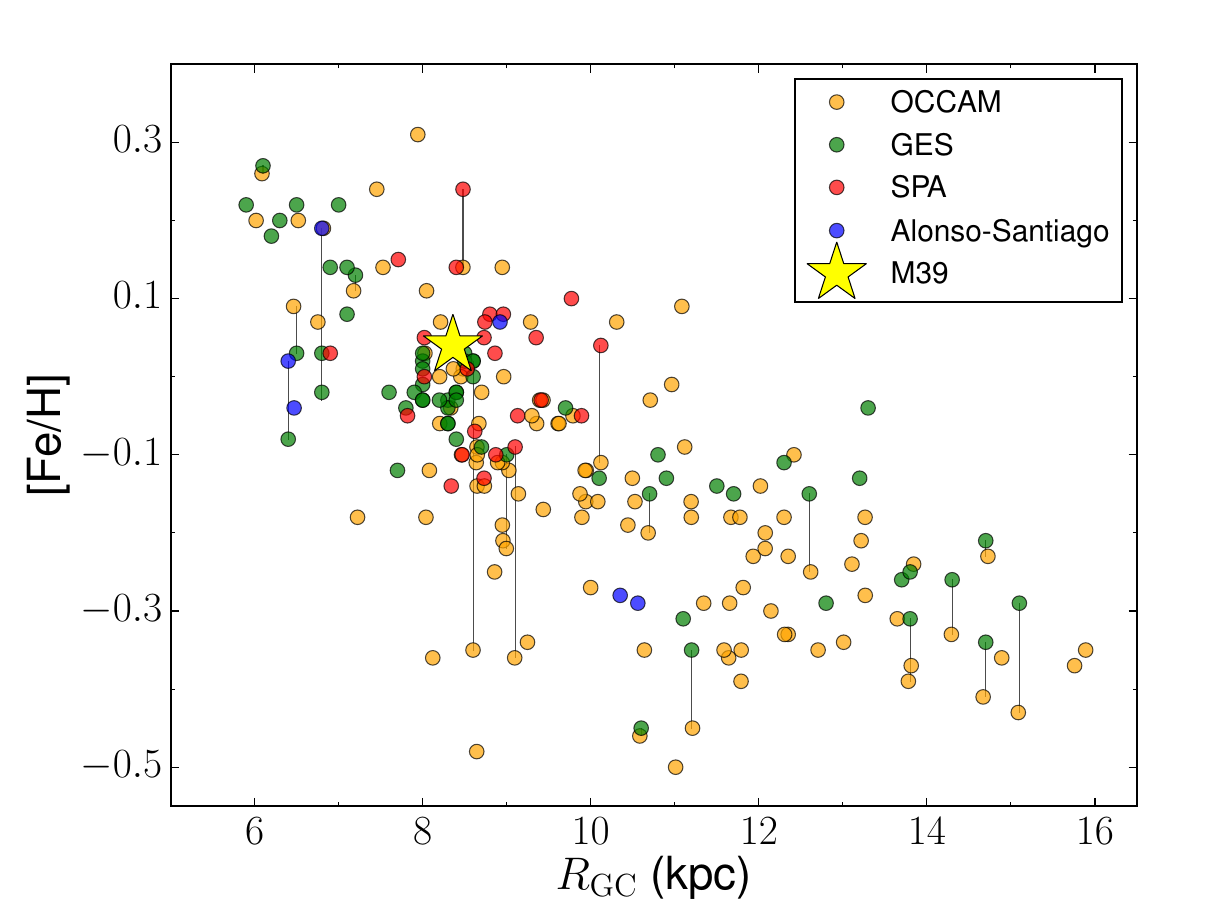}	
\caption{Radial metallicity gradient from open clusters studied in the framework of the $Gaia$-ESO and OCCAM Surveys, the SPA project and some other clusters analysed by 
Alonso-Santiago. Black lines link results for the same cluster provided by different authors. The star represents the metallicity found in this work for M\,39.}

\label{fig_grad}
\end{center}
\end{figure}

Open clusters are known to be among the best tracers of the radial distribution of metallicity in the Galaxy, the so-called Galactic gradient.
With the aim of evaluating how the metallicity obtained in this work for M\,39 compares with this gradient, we searched for a sample of homogeneously 
analysed clusters from the literature. We gathered the metallicity derived with high-resolution spectroscopy within the framework of 
the $Gaia$-ESO \citep{Randich22} and the OCCAM \citep[APOGEE-DR17,][]{Myers22} surveys. Additionally, clusters targeted in the SPA project 
\citep{ASCC123,D'Orazi20,Casali20,Zhang21,Stock2} and some other young open clusters studied by our group complete the sample \citep{6067,3105,2345,3OC}.
In total we collected a sample containing nearly two hundreds clusters at Galactocentric distances, $R_{\textrm{GC}}$, up to 16 kpc. 
The position of M\,39 on this gradient is shown in Fig.~\ref{fig_grad}. The metallicity, as the abundance of iron, was referenced to $A$(Fe)=7.45 \citep{Grevesse07}.
Galactocentric distances were taken from \citet{CG20}, who calculated them from $Gaia$-DR2 astrometry and taking as a reference a solar value of $R_{\odot}$=8.34\,kpc. 
As can be seen, the metallicity derived in this work is compatible with that expected for the cluster location.

\subsection{Chemical composition and Galactic trends}

\begin{figure*}[ht]
\begin{center}

\includegraphics[width=17cm]{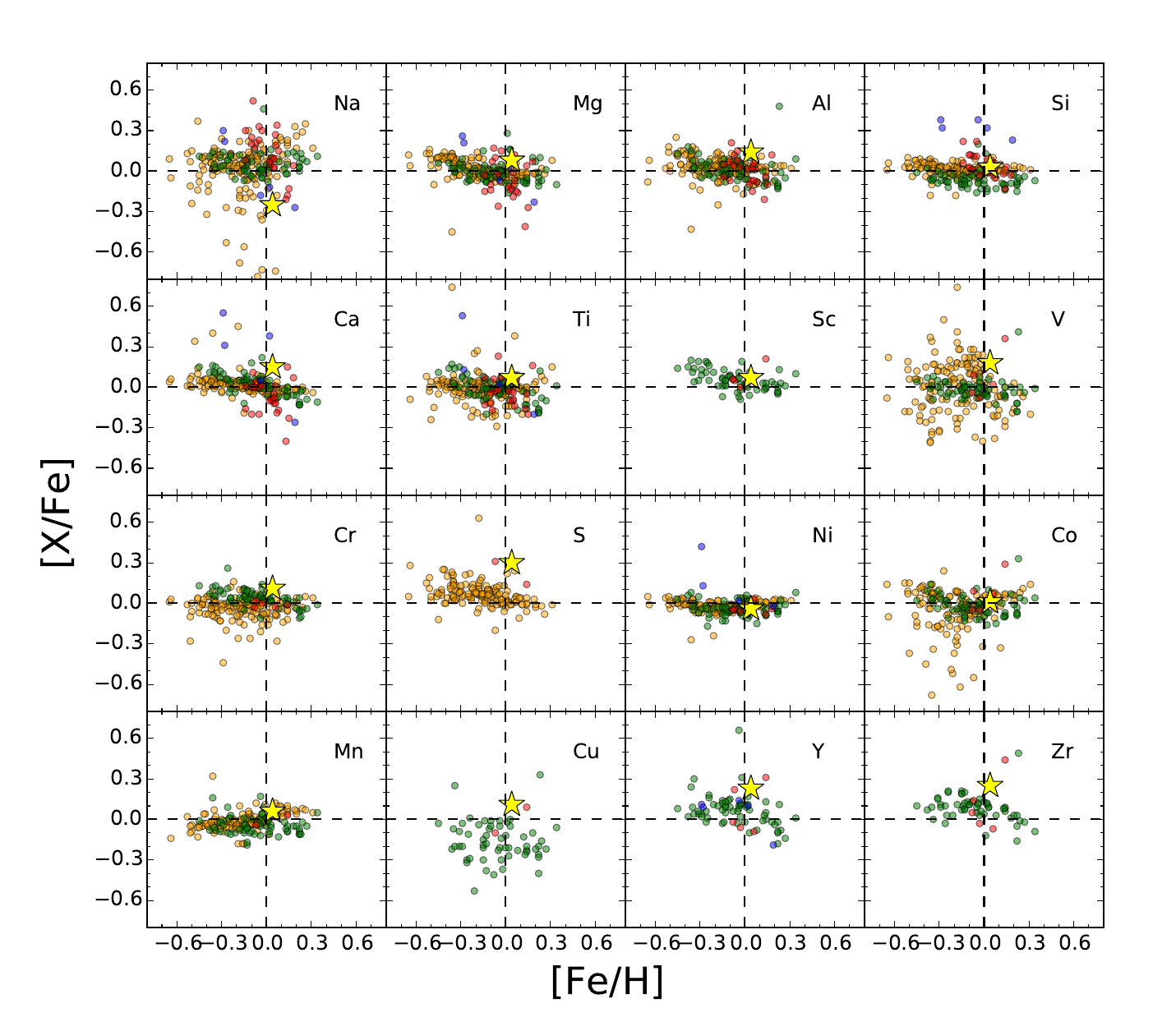}	
\caption{Abundance ratios [X/Fe] vs. [Fe/H]. Symbols and colours are the same as in Fig.~\ref{fig_grad}. The dashed lines show the solar value.}

\label{fig_trends}
\end{center}
\end{figure*}

As commented before (see Sect.~\ref{Sec:intro}) this work provides for the first time chemical abundances for M\,39. Thus, due to the lack of specific literature
with which to compare our results, we resorted to the chemical trends displayed by Galactic open clusters. We used the same sample previously employed when discussing 
about the Galactic gradient but in this case, chemical abundances for the GES (iDR6) and SPA clusters are taken from \citet{Magrini23} and \citet{Zhang22} respectively,
since \citet{Randich22} and \citet{Zhang21} only provided metallicity values. We compared the chemical composition of M\,39 with that of the other open clusters 
collected in our sample. Figure~\ref{fig_trends} shows the ratios [X/Fe] versus [Fe/H] for 16 chemical elements \citep[scaled to the solar abundances listed in][]{Grevesse07}.
The abundances of M\,39 are in a great agreement with the trends outlined by so many other open clusters collected in our sample and, therefore, we conclude that its 
chemical composition is fully compatible with that of the Galactic thin disc. All the elements are on the trends or sligthly above them. Only Na, the sole element in 
sample with a clearly subsolar abundance, and S are located further away from the bulk of the rest of the clusters, but always in a compatible place.

\section{Summary and conclusions}\label{Sec:concl}       
       
We conducted this research with the idea of adding one more tile to the mosaic of Galactic Archaeology that will help us improve our knowledge about the 
formation and evolution of the Milky Way. We focused on M\,39, a nearby young open cluster little-studied in recent years, whose chemical composition 
was so far unknown. We performed high-resolution spectroscopy with the HARPS-N and FIES spectrographs for 20 likely cluster members that were supplemented
with archival photometry and $Gaia$-DR3 data.

According to the literature, M\,39 is formed only by MS stars. In order to find any giant that belongs to the cluster we searched for new members. We recalculated 
the members identified by \citet{CG18} taking advantage of the latest $Gaia$-DR3 astrometry. We expand their list, finding 260 likely members within 
a radius of 250$\arcmin$ around the nominal cluster centre. Among them, no evolved stars have been found. By examining the spatial distribution of the members
we infer a distance to the cluster of 300\,pc.

We find three SB2s in our sample. For the remaining stars, we derived their radial and projected rotational velocities and estimated the extinction and
their atmospheric parameters. Additionally, for the coolest stars, nine in total, we carried out the chemical analysis, providing abundances for 21 elements. We 
do not find a significant reddening along the cluster field and estimate a mean RV of $-$5.5$\pm$0.5\,km\,s$^{-1}$. By fitting isochrones on three different CMDs
we find an age for the cluster of 430$\pm$110\,Ma, which is consistent with the content of Li and chromospheric activity found among the members.

M\,39 shows a solar-like metallicity, [Fe/H]=$+$0.04$\pm$0.08\,dex, which is consistent with that expected for its Galactocentric distance.
We performed the first detailed study of the chemical composition of the cluster to date. We determined the abundances of C, odd-Z elements (Na, Al), 
$\alpha$-elements (Mg, Si, S, Ca, Ti), Fe-peak elements (Sc, V, Cr, Mn, Co, Ni, Cu, Zn) and $s$-elements (Sr, Y, Zr, Ba). The cluster stars have a chemical 
composition similar to that of the Sun. Only Na shows a lower abundance, while S and the heaviest elements, specially Ba, display higher values.
Its ratios [X/Fe] are on the Galactic trends displayed by the large sample of open clusters studied in surveys such as the $Gaia$-ESO, OCCAM or SPA. 
The cluster shows solar-like mean ratios for $\alpha$ ([$\alpha$/Fe]=$+$0.12$\pm$0.10\,dex and Fe-peak elements ([X/Fe]=$+$0.05$\pm$0.10\,dex), while for the neutron-capture 
elements the ratio is slightly overabundant ([$s$/Fe]=$+$0.28$\pm$0.10\,dex). Finally, we conclude our research by claiming that the 
chemical composition of M\,39 is fully compatible with that of the Galactic thin disc.

\section*{Acknowledgements}

Based on observations made with the Italian \textit{Telescopio Nazionale Galileo} (TNG) operated on the island of La Palma by the Fundaci\'on Galileo
Galilei of the INAF (\textit{Istituto Nazionale di Astrofisica}) at the \textit{Observatorio del Roque de los Muchachos}.

This research is also partially based on observations made with the Nordic Optical Telescope, owned in collaboration by the University of Turku and Aarhus 
University, and operated jointly by Aarhus University, the University of Turku and the University of Oslo, representing Denmark, Finland and Norway, the 
University of Iceland and Stockholm University at the Observatorio del Roque de los Muchachos, La Palma, Spain, of the Instituto de Astrof\'isica de Canarias.

The research leading to these results has received funding from the European Union Seventh Framework Programme (F7/2007-2013) under grant agreement No. 
312430 (OPTICON).
 


\bibliographystyle{aa}
\bibliography{bib} 

\begin{thebibliography}{64}
\expandafter\ifx\csname natexlab\endcsname\relax\def\natexlab#1{#1}\fi

\bibitem[{{Abt} \& {Levato}(1976)}]{Abt76}
{Abt}, H.~A. \& {Levato}, H. 1976, \pasp, 88, 222

\bibitem[{{Abt} \& {Sanders}(1973)}]{Abt73}
{Abt}, H.~A. \& {Sanders}, W.~L. 1973, \apj, 186, 177

\bibitem[{{Allard}(2014)}]{Allard14}
{Allard}, F. 2014, in Exploring the Formation and Evolution of Planetary
  Systems, ed. M.~{Booth}, B.~C. {Matthews}, \& J.~R. {Graham}, Vol. 299,
  271--272

\bibitem[{{Alonso-Santiago} {et~al.}(2021){Alonso-Santiago}, {Frasca},
  {Catanzaro}, {Bragaglia}, {Andreuzzi}, {Carrera}, {Carretta}, {Casali},
  {D'Orazi}, {Fu}, {Giarrusso}, {Lucatello}, {Magrini}, {Origlia}, {Spina},
  {Vallenari}, \& {Zhang}}]{Stock2}
{Alonso-Santiago}, J., {Frasca}, A., {Catanzaro}, G., {et~al.} 2021, \aap, 656,
  A149

\bibitem[{{Alonso-Santiago} {et~al.}(2018){Alonso-Santiago}, {Marco},
  {Negueruela}, {Tabernero}, {Castro}, {McBride}, \& {Rajoelimanana}}]{3105}
{Alonso-Santiago}, J., {Marco}, A., {Negueruela}, I., {et~al.} 2018, \aap, 616,
  A124

\bibitem[{{Alonso-Santiago} {et~al.}(2020){Alonso-Santiago}, {Negueruela},
  {Marco}, {Tabernero}, \& {Castro}}]{3OC}
{Alonso-Santiago}, J., {Negueruela}, I., {Marco}, A., {Tabernero}, H.~M., \&
  {Castro}, N. 2020, \aap, 644, A136

\bibitem[{{Alonso-Santiago} {et~al.}(2017){Alonso-Santiago}, {Negueruela},
  {Marco}, {Tabernero}, {Gonz{\'a}lez-Fern{\'a}ndez}, \& {Castro}}]{6067}
{Alonso-Santiago}, J., {Negueruela}, I., {Marco}, A., {et~al.} 2017, \mnras,
  469, 1330

\bibitem[{{Alonso-Santiago} {et~al.}(2019){Alonso-Santiago}, {Negueruela},
  {Marco}, {Tabernero}, {Gonz{\'a}lez-Fern{\'a}ndez}, \& {Castro}}]{2345}
{Alonso-Santiago}, J., {Negueruela}, I., {Marco}, A., {et~al.} 2019, \aap, 631,
  A124

\bibitem[{{Andrae} {et~al.}(2023){Andrae}, {Fouesneau}, {Sordo},
  {Bailer-Jones}, {Dharmawardena}, {Rybizki}, {De Angeli}, {Lindstr{\o}m},
  {Marshall}, {Drimmel}, {Korn}, {Soubiran}, {Brouillet}, {Casamiquela}, {Rix},
  {Abreu Aramburu}, {{\'A}lvarez}, {Bakker}, {Bellas-Velidis}, {Bijaoui},
  {Brugaletta}, {Burlacu}, {Carballo}, {Chaoul}, {Chiavassa}, {Contursi},
  {Cooper}, {Creevey}, {Dafonte}, {Dapergolas}, {de Laverny}, {Delchambre},
  {Demouchy}, {Edvardsson}, {Fr{\'e}mat}, {Garabato}, {Garc{\'\i}a-Lario},
  {Garc{\'\i}a-Torres}, {Gavel}, {Gomez}, {Gonz{\'a}lez-Santamar{\'\i}a},
  {Hatzidimitriou}, {Heiter}, {Jean-Antoine Piccolo}, {Kontizas}, {Kordopatis},
  {Lanzafame}, {Lebreton}, {Licata}, {Livanou}, {Lobel}, {Lorca}, {Magdaleno
  Romeo}, {Manteiga}, {Marocco}, {Mary}, {Nicolas}, {Ordenovic}, {Pailler},
  {Palicio}, {Pallas-Quintela}, {Panem}, {Pichon}, {Poggio}, {Recio-Blanco},
  {Riclet}, {Robin}, {Santove{\~n}a}, {Sarro}, {Schultheis}, {Segol},
  {Silvelo}, {Slezak}, {Smart}, {S{\"u}veges}, {Th{\'e}venin}, {Torralba
  Elipe}, {Ulla}, {Utrilla}, {Vallenari}, {van Dillen}, {Zhao}, \&
  {Zorec}}]{Andrae23}
{Andrae}, R., {Fouesneau}, M., {Sordo}, R., {et~al.} 2023, \aap, 674, A27

\bibitem[{{Babusiaux} {et~al.}(2023){Babusiaux}, {Fabricius}, {Khanna},
  {Muraveva}, {Reyl{\'e}}, {Spoto}, {Vallenari}, {Luri}, {Arenou},
  {{\'A}lvarez}, {Anders}, {Antoja}, {Balbinot}, {Barache}, {Bauchet},
  {Bossini}, {Busonero}, {Cantat-Gaudin}, {Carrasco}, {Dafonte}, {Diakit{\'e}},
  {Figueras}, {Garcia-Gutierrez}, {Garofalo}, {Helmi}, {Jim{\'e}nez-Arranz},
  {Jordi}, {Kervella}, {Kostrzewa-Rutkowska}, {Leclerc}, {Licata}, {Manteiga},
  {Masip}, {Mongui{\'o}}, {Ramos}, {Robichon}, {Robin}, {Romero-G{\'o}mez},
  {S{\'a}ez}, {Santove{\~n}a}, {Spina}, {Torralba Elipe}, \&
  {Weiler}}]{Babusiaux23}
{Babusiaux}, C., {Fabricius}, C., {Khanna}, S., {et~al.} 2023, \aap, 674, A32

\bibitem[{{Cantat-Gaudin} {et~al.}(2020){Cantat-Gaudin}, {Anders},
  {Castro-Ginard}, {Jordi}, {Romero-G{\'o}mez}, {Soubiran}, {Casamiquela},
  {Tarricq}, {Moitinho}, {Vallenari}, {Bragaglia}, {Krone-Martins}, \&
  {Kounkel}}]{CG20}
{Cantat-Gaudin}, T., {Anders}, F., {Castro-Ginard}, A., {et~al.} 2020, \aap,
  640, A1

\bibitem[{{Cantat-Gaudin} {et~al.}(2018){Cantat-Gaudin}, {Jordi}, {Vallenari},
  {Bragaglia}, {Balaguer-N{\'u}{\~n}ez}, {Soubiran}, {Bossini}, {Moitinho},
  {Castro-Ginard}, {Krone-Martins}, {Casamiquela}, {Sordo}, \&
  {Carrera}}]{CG18}
{Cantat-Gaudin}, T., {Jordi}, C., {Vallenari}, A., {et~al.} 2018, \aap, 618,
  A93

\bibitem[{{Casali} {et~al.}(2020){Casali}, {Magrini}, {Frasca}, {Bragaglia},
  {Catanzaro}, {D'Orazi}, {Sordo}, {Carretta}, {Origlia}, {Andreuzzi}, {Fu}, \&
  {Vallenari}}]{Casali20}
{Casali}, G., {Magrini}, L., {Frasca}, A., {et~al.} 2020, \aap, 643, A12

\bibitem[{{Catanzaro} {et~al.}(2011){Catanzaro}, {Ripepi}, {Bernabei},
  {Marconi}, {Balona}, {Kurtz}, {Smalley}, {Borucki}, {Bruntt},
  {Christensen-Dalsgaard}, {Grigahc{\`e}ne}, {Kjeldsen}, {Koch}, {Monteiro},
  {Su{\'a}rez}, {Szab{\'o}}, \& {Uytterhoeven}}]{Catanzaro11}
{Catanzaro}, G., {Ripepi}, V., {Bernabei}, S., {et~al.} 2011, \mnras, 411, 1167

\bibitem[{{Catanzaro} {et~al.}(2013){Catanzaro}, {Ripepi}, \&
  {Bruntt}}]{Catanzaro13}
{Catanzaro}, G., {Ripepi}, V., \& {Bruntt}, H. 2013, \mnras, 431, 3258

\bibitem[{{Cosentino} {et~al.}(2014){Cosentino}, {Lovis}, {Pepe}, {Collier
  Cameron}, {Latham}, {Molinari}, {Udry}, {Bezawada}, {Buchschacher},
  {Figueira}, {Fleury}, {Ghedina}, {Glenday}, {Gonzalez}, {Guerra}, {Henry},
  {Hughes}, {Maire}, {Motalebi}, \& {Phillips}}]{Cosentino14}
{Cosentino}, R., {Lovis}, C., {Pepe}, F., {et~al.} 2014, in Society of
  Photo-Optical Instrumentation Engineers (SPIE) Conference Series, Vol. 9147,
  Ground-based and Airborne Instrumentation for Astronomy V, ed. S.~K.
  {Ramsay}, I.~S. {McLean}, \& H.~{Takami}, 91478C

\bibitem[{{Cutri} \& {et al.}(2012)}]{WISE}
{Cutri}, R.~M. \& {et al.} 2012, VizieR Online Data Catalog, II/311

\bibitem[{{de Jong} {et~al.}(2019){de Jong}, {Agertz}, {Berbel}, {Aird},
  {Alexander}, {Amarsi}, {Anders}, {Andrae}, {Ansarinejad}, {Ansorge},
  {Antilogus}, {Anwand-Heerwart}, {Arentsen}, {Arnadottir}, {Asplund}, {Auger},
  {Azais}, {Baade}, {Baker}, {Baker}, {Balbinot}, {Baldry}, {Banerji},
  {Barden}, {Barklem}, {Barth{\'e}l{\'e}my-Mazot}, {Battistini}, {Bauer},
  {Bell}, {Bellido-Tirado}, {Bellstedt}, {Belokurov}, {Bensby}, {Bergemann},
  {Bestenlehner}, {Bielby}, {Bilicki}, {Blake}, {Bland-Hawthorn}, {Boeche},
  {Boland}, {Boller}, {Bongard}, {Bongiorno}, {Bonifacio}, {Boudon}, {Brooks},
  {Brown}, {Brown}, {Br{\"u}ggen}, {Brynnel}, {Brzeski}, {Buchert},
  {Buschkamp}, {Caffau}, {Caillier}, {Carrick}, {Casagrande}, {Case}, {Casey},
  {Cesarini}, {Cescutti}, {Chapuis}, {Chiappini}, {Childress}, {Christlieb},
  {Church}, {Cioni}, {Cluver}, {Colless}, {Collett}, {Comparat}, {Cooper},
  {Couch}, {Courbin}, {Croom}, {Croton}, {Daguis{\'e}}, {Dalton}, {Davies},
  {Davis}, {de Laverny}, {Deason}, {Dionies}, {Disseau}, {Doel}, {D{\"o}scher},
  {Driver}, {Dwelly}, {Eckert}, {Edge}, {Edvardsson}, {Youssoufi}, {Elhaddad},
  {Enke}, {Erfanianfar}, {Farrell}, {Fechner}, {Feiz}, {Feltzing}, {Ferreras},
  {Feuerstein}, {Feuillet}, {Finoguenov}, {Ford}, {Fotopoulou}, {Fouesneau},
  {Frenk}, {Frey}, {Gaessler}, {Geier}, {Gentile Fusillo}, {Gerhard},
  {Giannantonio}, {Giannone}, {Gibson}, {Gillingham},
  {Gonz{\'a}lez-Fern{\'a}ndez}, {Gonzalez-Solares}, {Gottloeber}, {Gould},
  {Grebel}, {Gueguen}, {Guiglion}, {Haehnelt}, {Hahn}, {Hansen}, {Hartman},
  {Hauptner}, {Hawkins}, {Haynes}, {Haynes}, {Heiter}, {Helmi}, {Aguayo},
  {Hewett}, {Hinton}, {Hobbs}, {Hoenig}, {Hofman}, {Hook}, {Hopgood},
  {Hopkins}, {Hourihane}, {Howes}, {Howlett}, {Huet}, {Irwin}, {Iwert},
  {Jablonka}, {Jahn}, {Jahnke}, {Jarno}, {Jin}, {Jofre}, {Johl}, {Jones},
  {J{\"o}nsson}, {Jordan}, {Karovicova}, {Khalatyan}, {Kelz}, {Kennicutt},
  {King}, {Kitaura}, {Klar}, {Klauser}, {Kneib}, {Koch}, {Koposov},
  {Kordopatis}, {Korn}, {Kosmalski}, {Kotak}, {Kovalev}, {Kreckel}, {Kripak},
  {Krumpe}, {Kuijken}, {Kunder}, {Kushniruk}, {Lam}, {Lamer}, {Laurent},
  {Lawrence}, {Lehmitz}, {Lemasle}, {Lewis}, {Li}, {Lidman}, {Lind}, {Liske},
  {Lizon}, {Loveday}, {Ludwig}, {McDermid}, {Maguire}, {Mainieri}, {Mali},
  {Mandel}, {Mandel}, {Mannering}, {Martell}, {Martinez Delgado}, {Matijevic},
  {McGregor}, {McMahon}, {McMillan}, {Mena}, {Merloni}, {Meyer}, {Michel},
  {Micheva}, {Migniau}, {Minchev}, {Monari}, {Muller}, {Murphy},
  {Muthukrishna}, {Nandra}, {Navarro}, {Ness}, {Nichani}, {Nichol}, {Nicklas},
  {Niederhofer}, {Norberg}, {Obreschkow}, {Oliver}, {Owers}, {Pai},
  {Pankratow}, {Parkinson}, {Paschke}, {Paterson}, {Pecontal}, {Parry},
  {Phillips}, {Pillepich}, {Pinard}, {Pirard}, {Piskunov}, {Plank},
  {Pl{\"u}schke}, {Pons}, {Popesso}, {Power}, {Pragt}, {Pramskiy}, {Pryer},
  {Quattri}, {Queiroz}, {Quirrenbach}, {Rahurkar}, {Raichoor}, {Ramstedt},
  {Rau}, {Recio-Blanco}, {Reiss}, {Renaud}, {Revaz}, {Rhode}, {Richard},
  {Richter}, {Rix}, {Robotham}, {Roelfsema}, {Romaniello}, {Rosario},
  {Rothmaier}, {Roukema}, {Ruchti}, {Rupprecht}, {Rybizki}, {Ryde}, {Saar},
  {Sadler}, {Sahl{\'e}n}, {Salvato}, {Sassolas}, {Saunders}, {Saviauk},
  {Sbordone}, {Schmidt}, {Schnurr}, {Scholz}, {Schwope}, {Seifert}, {Shanks},
  {Sheinis}, {Sivov}, {Sk{\'u}lad{\'o}ttir}, {Smartt}, {Smedley}, {Smith},
  {Smith}, {Sorce}, {Spitler}, {Starkenburg}, {Steinmetz}, {Stilz}, {Storm},
  {Sullivan}, {Sutherland}, {Swann}, {Tamone}, {Taylor}, {Teillon}, {Tempel},
  {ter Horst}, {Thi}, {Tolstoy}, {Trager}, {Traven}, {Tremblay}, {Tresse},
  {Valentini}, {van de Weygaert}, {van den Ancker}, {Veljanoski}, {Venkatesan},
  {Wagner}, {Wagner}, {Walcher}, {Waller}, {Walton}, {Wang}, {Winkler},
  {Wisotzki}, {Worley}, {Worseck}, {Xiang}, {Xu}, {Yong}, {Zhao}, {Zheng},
  {Zscheyge}, \& {Zucker}}]{4MOST}
{de Jong}, R.~S., {Agertz}, O., {Berbel}, A.~A., {et~al.} 2019, The Messenger,
  175, 3

\bibitem[{{De Silva} {et~al.}(2015){De Silva}, {Freeman}, {Bland-Hawthorn},
  {Martell}, {de Boer}, {Asplund}, {Keller}, {Sharma}, {Zucker}, {Zwitter},
  {Anguiano}, {Bacigalupo}, {Bayliss}, {Beavis}, {Bergemann}, {Campbell},
  {Cannon}, {Carollo}, {Casagrande}, {Casey}, {Da Costa}, {D'Orazi}, {Dotter},
  {Duong}, {Heger}, {Ireland}, {Kafle}, {Kos}, {Lattanzio}, {Lewis}, {Lin},
  {Lind}, {Munari}, {Nataf}, {O'Toole}, {Parker}, {Reid}, {Schlesinger},
  {Sheinis}, {Simpson}, {Stello}, {Ting}, {Traven}, {Watson}, {Wittenmyer},
  {Yong}, \& {{\v{Z}}erjal}}]{GALAH}
{De Silva}, G.~M., {Freeman}, K.~C., {Bland-Hawthorn}, J., {et~al.} 2015,
  \mnras, 449, 2604

\bibitem[{{D'Orazi} {et~al.}(2020){D'Orazi}, {Oliva}, {Bragaglia}, {Frasca},
  {Sanna}, {Biazzo}, {Casali}, {Desidera}, {Lucatello}, {Magrini}, \&
  {Origlia}}]{D'Orazi20}
{D'Orazi}, V., {Oliva}, E., {Bragaglia}, A., {et~al.} 2020, \aap, 633, A38

\bibitem[{{Ebbighausen}(1940)}]{Ebbighausen40}
{Ebbighausen}, E.~G. 1940, \apj, 92, 434

\bibitem[{{Eggen}(1951)}]{Eggen51}
{Eggen}, O.~J. 1951, \apj, 113, 657

\bibitem[{{Fouesneau} {et~al.}(2023){Fouesneau}, {Fr{\'e}mat}, {Andrae},
  {Korn}, {Soubiran}, {Kordopatis}, {Vallenari}, {Heiter}, {Creevey}, {Sarro},
  {de Laverny}, {Lanzafame}, {Lobel}, {Sordo}, {Rybizki}, {Slezak},
  {{\'A}lvarez}, {Drimmel}, {Garabato}, {Delchambre}, {Bailer-Jones},
  {Hatzidimitriou}, {Lorca}, {Le Fustec}, {Pailler}, {Mary}, {Robin},
  {Utrilla}, {Abreu Aramburu}, {Bakker}, {Bellas-Velidis}, {Bijaoui}, {Blomme},
  {Bouret}, {Brouillet}, {Brugaletta}, {Burlacu}, {Carballo}, {Casamiquela},
  {Chaoul}, {Chiavassa}, {Contursi}, {Cooper}, {Dafonte}, {Demouchy},
  {Dharmawardena}, {Garc{\'\i}a-Lario}, {Garc{\'\i}a-Torres}, {Gomez},
  {Gonz{\'a}lez-Santamar{\'\i}a}, {Jean-Antoine Piccolo}, {Kontizas},
  {Lebreton}, {Licata}, {Lindstr{\o}m}, {Livanou}, {Magdaleno Romeo},
  {Manteiga}, {Marocco}, {Martayan}, {Marshall}, {Nicolas}, {Ordenovic},
  {Palicio}, {Pallas-Quintela}, {Pichon}, {Poggio}, {Recio-Blanco}, {Riclet},
  {Santove{\~n}a}, {Schultheis}, {Segol}, {Silvelo}, {Smart}, {S{\"u}veges},
  {Th{\'e}venin}, {Torralba Elipe}, {Ulla}, {van Dillen}, {Zhao}, \&
  {Zorec}}]{Fouesneau23}
{Fouesneau}, M., {Fr{\'e}mat}, Y., {Andrae}, R., {et~al.} 2023, \aap, 674, A28

\bibitem[{{Frasca} {et~al.}(2023{\natexlab{a}}){Frasca}, {Alonso-Santiago},
  {Catanzaro}, \& {Bragaglia}}]{Frasca23a}
{Frasca}, A., {Alonso-Santiago}, J., {Catanzaro}, G., \& {Bragaglia}, A.
  2023{\natexlab{a}}, \mnras, 522, 4894

\bibitem[{{Frasca} {et~al.}(2019){Frasca}, {Alonso-Santiago}, {Catanzaro},
  {Bragaglia}, {Carretta}, {Casali}, {D'Orazi}, {Magrini}, {Andreuzzi},
  {Oliva}, {Origlia}, {Sordo}, \& {Vallenari}}]{ASCC123}
{Frasca}, A., {Alonso-Santiago}, J., {Catanzaro}, G., {et~al.} 2019, \aap, 632,
  A16

\bibitem[{{Frasca} {et~al.}(2023{\natexlab{b}}){Frasca}, {Alonso-Santiago},
  {Catanzaro}, {Bragaglia}, {D'Orazi}, {Fu}, {Vallenari}, \&
  {Andreuzzi}}]{Frasca23b}
{Frasca}, A., {Alonso-Santiago}, J., {Catanzaro}, G., {et~al.}
  2023{\natexlab{b}}, arXiv e-prints, arXiv:2307.14081

\bibitem[{{Frasca} {et~al.}(2015){Frasca}, {Biazzo}, {Lanzafame}, {Alcal{\'a}},
  {Brugaletta}, {Klutsch}, {Stelzer}, {Sacco}, {Spina}, {Jeffries}, {Montes},
  {Alfaro}, {Barentsen}, {Bonito}, {Gameiro}, {L{\'o}pez-Santiago}, {Pace},
  {Pasquini}, {Prisinzano}, {Sousa}, {Gilmore}, {Randich}, {Micela},
  {Bragaglia}, {Flaccomio}, {Bayo}, {Costado}, {Franciosini}, {Hill},
  {Hourihane}, {Jofr{\'e}}, {Lardo}, {Maiorca}, {Masseron}, {Morbidelli}, \&
  {Worley}}]{Frasca15}
{Frasca}, A., {Biazzo}, K., {Lanzafame}, A.~C., {et~al.} 2015, \aap, 575, A4

\bibitem[{{Frasca} {et~al.}(2000){Frasca}, {Freire Ferrero}, {Marilli}, \&
  {Catalano}}]{Frasca2000}
{Frasca}, A., {Freire Ferrero}, R., {Marilli}, E., \& {Catalano}, S. 2000,
  \aap, 364, 179

\bibitem[{{Frasca} {et~al.}(2018){Frasca}, {Guillout}, {Klutsch}, {Ferrero},
  {Marilli}, {Biazzo}, {Gandolfi}, \& {Montes}}]{Frasca18}
{Frasca}, A., {Guillout}, P., {Klutsch}, A., {et~al.} 2018, \aap, 612, A96

\bibitem[{{Frasca} {et~al.}(2006){Frasca}, {Guillout}, {Marilli}, {Freire
  Ferrero}, {Biazzo}, \& {Klutsch}}]{Frasca06}
{Frasca}, A., {Guillout}, P., {Marilli}, E., {et~al.} 2006, \aap, 454, 301

\bibitem[{{Gaia Collaboration} {et~al.}(2022){Gaia Collaboration}, {Vallenari},
  {Brown}, {Prusti}, {de Bruijne}, {Arenou}, {Babusiaux}, {Biermann},
  {Creevey}, {Ducourant}, {Evans}, {Eyer}, {Guerra}, {Hutton}, {Jordi},
  {Klioner}, {Lammers}, {Lindegren}, {Luri}, {Mignard}, {Panem}, {Pourbaix},
  {Randich}, {Sartoretti}, {Soubiran}, {Tanga}, {Walton}, {Bailer-Jones},
  {Bastian}, {Drimmel}, {Jansen}, {Katz}, {Lattanzi}, {van Leeuwen}, {Bakker},
  {Cacciari}, {Casta{\~n}eda}, {De Angeli}, {Fabricius}, {Fouesneau},
  {Fr{\'e}mat}, {Galluccio}, {Guerrier}, {Heiter}, {Masana}, {Messineo},
  {Mowlavi}, {Nicolas}, {Nienartowicz}, {Pailler}, {Panuzzo}, {Riclet}, {Roux},
  {Seabroke}, {Sordo{\o}rcit}, {Th{\'e}venin}, {Gracia-Abril}, {Portell},
  {Teyssier}, {Altmann}, {Andrae}, {Audard}, {Bellas-Velidis}, {Benson},
  {Berthier}, {Blomme}, {Burgess}, {Busonero}, {Busso}, {C{\'a}novas}, {Carry},
  {Cellino}, {Cheek}, {Clementini}, {Damerdji}, {Davidson}, {de Teodoro},
  {Nu{\~n}ez Campos}, {Delchambre}, {Dell'Oro}, {Esquej},
  {Fern{\'a}ndez-Hern{\'a}ndez}, {Fraile}, {Garabato}, {Garc{\'\i}a-Lario},
  {Gosset}, {Haigron}, {Halbwachs}, {Hambly}, {Harrison}, {Hern{\'a}ndez},
  {Hestroffer}, {Hodgkin}, {Holl}, {Jan{\ss}en}, {Jevardat de Fombelle},
  {Jordan}, {Krone-Martins}, {Lanzafame}, {L{\"o}ffler}, {Marchal}, {Marrese},
  {Moitinho}, {Muinonen}, {Osborne}, {Pancino}, {Pauwels}, {Recio-Blanco},
  {Reyl{\'e}}, {Riello}, {Rimoldini}, {Roegiers}, {Rybizki}, {Sarro}, {Siopis},
  {Smith}, {Sozzetti}, {Utrilla}, {van Leeuwen}, {Abbas}, {{\'A}brah{\'a}m},
  {Abreu Aramburu}, {Aerts}, {Aguado}, {Ajaj}, {Aldea-Montero}, {Altavilla},
  {{\'A}lvarez}, {Alves}, {Anders}, {Anderson}, {Anglada Varela}, {Antoja},
  {Baines}, {Baker}, {Balaguer-N{\'u}{\~n}ez}, {Balbinot}, {Balog}, {Barache},
  {Barbato}, {Barros}, {Barstow}, {Bartolom{\'e}}, {Bassilana}, {Bauchet},
  {Becciani}, {Bellazzini}, {Berihuete}, {Bernet}, {Bertone}, {Bianchi},
  {Binnenfeld}, {Blanco-Cuaresma}, {Blazere}, {Boch}, {Bombrun}, {Bossini},
  {Bouquillon}, {Bragaglia}, {Bramante}, {Breedt}, {Bressan}, {Brouillet},
  {Brugaletta}, {Bucciarelli}, {Burlacu}, {Butkevich}, {Buzzi}, {Caffau},
  {Cancelliere}, {Cantat-Gaudin}, {Carballo}, {Carlucci}, {Carnerero},
  {Carrasco}, {Casamiquela}, {Castellani}, {Castro-Ginard}, {Chaoul},
  {Charlot}, {Chemin}, {Chiaramida}, {Chiavassa}, {Chornay}, {Comoretto},
  {Contursi}, {Cooper}, {Cornez}, {Cowell}, {Crifo}, {Cropper}, {Crosta},
  {Crowley}, {Dafonte}, {Dapergolas}, {David}, {David}, {de Laverny}, {De
  Luise}, {De March}, {De Ridder}, {de Souza}, {de Torres}, {del Peloso}, {del
  Pozo}, {Delbo}, {Delgado}, {Delisle}, {Demouchy}, {Dharmawardena}, {Di
  Matteo}, {Diakite}, {Diener}, {Distefano}, {Dolding}, {Edvardsson}, {Enke},
  {Fabre}, {Fabrizio}, {Faigler}, {Fedorets}, {Fernique}, {Fienga}, {Figueras},
  {Fournier}, {Fouron}, {Fragkoudi}, {Gai}, {Garcia-Gutierrez},
  {Garcia-Reinaldos}, {Garc{\'\i}a-Torres}, {Garofalo}, {Gavel}, {Gavras},
  {Gerlach}, {Geyer}, {Giacobbe}, {Gilmore}, {Girona}, {Giuffrida}, {Gomel},
  {Gomez}, {Gonz{\'a}lez-N{\'u}{\~n}ez}, {Gonz{\'a}lez-Santamar{\'\i}a},
  {Gonz{\'a}lez-Vidal}, {Granvik}, {Guillout}, {Guiraud},
  {Guti{\'e}rrez-S{\'a}nchez}, {Guy}, {Hatzidimitriou}, {Hauser}, {Haywood},
  {Helmer}, {Helmi}, {Sarmiento}, {Hidalgo}, {Hilger}, {H{\l}adczuk}, {Hobbs},
  {Holland}, {Huckle}, {Jardine}, {Jasniewicz}, {Jean-Antoine Piccolo},
  {Jim{\'e}nez-Arranz}, {Jorissen}, {Juaristi Campillo}, {Julbe}, {Karbevska},
  {Kervella}, {Khanna}, {Kontizas}, {Kordopatis}, {Korn}, {K{\'o}sp{\'a}l},
  {Kostrzewa-Rutkowska}, {Kruszy{\'n}ska}, {Kun}, {Laizeau}, {Lambert},
  {Lanza}, {Lasne}, {Le Campion}, {Lebreton}, {Lebzelter}, {Leccia}, {Leclerc},
  {Lecoeur-Taibi}, {Liao}, {Licata}, {Lindstr{\o}m}, {Lister}, {Livanou},
  {Lobel}, {Lorca}, {Loup}, {Madrero Pardo}, {Magdaleno Romeo}, {Managau},
  {Mann}, {Manteiga}, {Marchant}, {Marconi}, {Marcos}, {Marcos Santos},
  {Mar{\'\i}n Pina}, {Marinoni}, {Marocco}, {Marshall}, {Polo},
  {Mart{\'\i}n-Fleitas}, {Marton}, {Mary}, {Masip}, {Massari},
  {Mastrobuono-Battisti}, {Mazeh}, {McMillan}, {Messina}, {Michalik}, {Millar},
  {Mints}, {Molina}, {Molinaro}, {Moln{\'a}r}, {Monari}, {Mongui{\'o}},
  {Montegriffo}, {Montero}, {Mor}, {Mora}, {Morbidelli}, {Morel}, {Morris},
  {Muraveva}, {Murphy}, {Musella}, {Nagy}, {Noval}, {Oca{\~n}a}, {Ogden},
  {Ordenovic}, {Osinde}, {Pagani}, {Pagano}, {Palaversa}, {Palicio},
  {Pallas-Quintela}, {Panahi}, {Payne-Wardenaar}, {Pe{\~n}alosa Esteller},
  {Penttil{\"a}}, {Pichon}, {Piersimoni}, {Pineau}, {Plachy}, {Plum}, {Poggio},
  {Pr{\v{s}}a}, {Pulone}, {Racero}, {Ragaini}, {Rainer}, {Raiteri}, {Rambaux},
  {Ramos}, {Ramos-Lerate}, {Re Fiorentin}, {Regibo}, {Richards}, {Rios Diaz},
  {Ripepi}, {Riva}, {Rix}, {Rixon}, {Robichon}, {Robin}, {Robin}, {Roelens},
  {Rogues}, {Rohrbasser}, {Romero-G{\'o}mez}, {Rowell}, {Royer}, {Ruz Mieres},
  {Rybicki}, {Sadowski}, {S{\'a}ez N{\'u}{\~n}ez}, {Sagrist{\`a} Sell{\'e}s},
  {Sahlmann}, {Salguero}, {Samaras}, {Sanchez Gimenez}, {Sanna},
  {Santove{\~n}a}, {Sarasso}, {Schultheis}, {Sciacca}, {Segol}, {Segovia},
  {S{\'e}gransan}, {Semeux}, {Shahaf}, {Siddiqui}, {Siebert}, {Siltala},
  {Silvelo}, {Slezak}, {Slezak}, {Smart}, {Snaith}, {Solano}, {Solitro},
  {Souami}, {Souchay}, {Spagna}, {Spina}, {Spoto}, {Steele},
  {Steidelm{\"u}ller}, {Stephenson}, {S{\"u}veges}, {Surdej}, {Szabados},
  {Szegedi-Elek}, {Taris}, {Taylo}, {Teixeira}, {Tolomei}, {Tonello}, {Torra},
  {Torra}, {Torralba Elipe}, {Trabucchi}, {Tsounis}, {Turon}, {Ulla}, {Unger},
  {Vaillant}, {van Dillen}, {van Reeven}, {Vanel}, {Vecchiato}, {Viala},
  {Vicente}, {Voutsinas}, {Weiler}, {Wevers}, {Wyrzykowski}, {Yoldas}, {Yvard},
  {Zhao}, {Zorec}, {Zucker}, \& {Zwitter}}]{DR3}
{Gaia Collaboration}, {Vallenari}, A., {Brown}, A.~G.~A., {et~al.} 2022, arXiv
  e-prints, arXiv:2208.00211

\bibitem[{{Gilmore} {et~al.}(2022){Gilmore}, {Randich}, {Worley}, {Hourihane},
  {Gonneau}, {Sacco}, {Lewis}, {Magrini}, {Fran{\c{c}}ois}, {Jeffries},
  {Koposov}, {Bragaglia}, {Alfaro}, {Allende Prieto}, {Blomme}, {Korn},
  {Lanzafame}, {Pancino}, {Recio-Blanco}, {Smiljanic}, {Van Eck}, {Zwitter},
  {Bensby}, {Flaccomio}, {Irwin}, {Franciosini}, {Morbidelli}, {Damiani},
  {Bonito}, {Friel}, {Vink}, {Prisinzano}, {Abbas}, {Hatzidimitriou}, {Held},
  {Jordi}, {Paunzen}, {Spagna}, {Jackson}, {Ma{\'\i}z Apell{\'a}niz},
  {Asplund}, {Bonifacio}, {Feltzing}, {Binney}, {Drew}, {Ferguson}, {Micela},
  {Negueruela}, {Prusti}, {Rix}, {Vallenari}, {Bergemann}, {Casey}, {de
  Laverny}, {Frasca}, {Hill}, {Lind}, {Sbordone}, {Sousa}, {Adibekyan},
  {Caffau}, {Daflon}, {Feuillet}, {Gebran}, {Gonzalez Hernandez}, {Guiglion},
  {Herrero}, {Lobel}, {Merle}, {Mikolaitis}, {Montes}, {Morel}, {Ruchti},
  {Soubiran}, {Tabernero}, {Tautvai{\v{s}}ien{\.{e}}}, {Traven}, {Valentini},
  {Van der Swaelmen}, {Villanova}, {Viscasillas V{\'a}zquez}, {Bayo}, {Biazzo},
  {Carraro}, {Edvardsson}, {Heiter}, {Jofr{\'e}}, {Marconi}, {Martayan},
  {Masseron}, {Monaco}, {Walton}, {Zaggia}, {Aguirre B{\o}rsen-Koch}, {Alves},
  {Balaguer-Nunez}, {Barklem}, {Barrado}, {Bellazzini}, {Berlanas}, {Binks},
  {Bressan}, {Capuzzo-Dolcetta}, {Casagrande}, {Casamiquela}, {Collins},
  {D'Orazi}, {Dantas}, {Debattista}, {Delgado-Mena}, {Di Marcantonio},
  {Drazdauskas}, {Evans}, {Famaey}, {Franchini}, {Fr{\'e}mat}, {Fu}, {Geisler},
  {Gerhard}, {Gonz{\'a}lez Solares}, {Grebel}, {Guti{\'e}rrez Albarr{\'a}n},
  {Jim{\'e}nez-Esteban}, {J{\"o}nsson}, {Khachaturyants}, {Kordopatis}, {Kos},
  {Lagarde}, {Ludwig}, {Mahy}, {Mapelli}, {Marfil}, {Martell}, {Messina},
  {Miglio}, {Minchev}, {Moitinho}, {Montalban}, {Monteiro}, {Morossi},
  {Mowlavi}, {Mucciarelli}, {Murphy}, {Nardetto}, {Ortolani}, {Paletou},
  {Palou{\v{s}}}, {Pickering}, {Quirrenbach}, {Re Fiorentin}, {Read}, {Romano},
  {Ryde}, {Sanna}, {Santos}, {Seabroke}, {Spina}, {Steinmetz}, {Stonkut{\'e}},
  {Sutorius}, {Th{\'e}venin}, {Tosi}, {Tsantaki}, {Wright}, {Wyse}, {Zoccali},
  {Zorec}, \& {Zucker}}]{Gilmore22}
{Gilmore}, G., {Randich}, S., {Worley}, C.~C., {et~al.} 2022, \aap, 666, A120

\bibitem[{{Grevesse} {et~al.}(2007){Grevesse}, {Asplund}, \&
  {Sauval}}]{Grevesse07}
{Grevesse}, N., {Asplund}, M., \& {Sauval}, A.~J. 2007, \ssr, 130, 105

\bibitem[{{Henden} {et~al.}(2016){Henden}, {Templeton}, {Terrell}, {Smith},
  {Levine}, \& {Welch}}]{APASS}
{Henden}, A.~A., {Templeton}, M., {Terrell}, D., {et~al.} 2016, VizieR Online
  Data Catalog, II/336

\bibitem[{{Hunt} \& {Reffert}(2023)}]{Hunt23}
{Hunt}, E.~L. \& {Reffert}, S. 2023, \aap, 673, A114

\bibitem[{{Jeffries}(2014)}]{Jeffries14}
{Jeffries}, R.~D. 2014, in EAS Publications Series, Vol.~65, EAS Publications
  Series, ed. Y.~{Lebreton}, D.~{Valls-Gabaud}, \& C.~{Charbonnel}, 289--325

\bibitem[{{Jeffries} {et~al.}(2023){Jeffries}, {Jackson}, {Wright}, {Weaver},
  {Gilmore}, {Randich}, {Bragaglia}, {Korn}, {Smiljanic}, {Biazzo}, {Casey},
  {Frasca}, {Gonneau}, {Guiglion}, {Morbidelli}, {Prisinzano}, {Sacco},
  {Tautvai{\v{s}}ien{\.{e}}}, {Worley}, \& {Zaggia}}]{Jeffries23}
{Jeffries}, R.~D., {Jackson}, R.~J., {Wright}, N.~J., {et~al.} 2023, \mnras,
  523, 802

\bibitem[{{Jin} {et~al.}(2023){Jin}, {Trager}, {Dalton}, {Aguerri}, {Drew},
  {Falc{\'o}n-Barroso}, {G{\"a}nsicke}, {Hill}, {Iovino}, {Pieri}, {Poggianti},
  {Smith}, {Vallenari}, {Abrams}, {Aguado}, {Antoja}, {Arag{\'o}n-Salamanca},
  {Ascasibar}, {Babusiaux}, {Balcells}, {Barrena}, {Battaglia}, {Belokurov},
  {Bensby}, {Bonifacio}, {Bragaglia}, {Carrasco}, {Carrera}, {Cornwell},
  {Dom{\'\i}nguez-Palmero}, {Duncan}, {Famaey}, {Fari{\~n}a}, {Gonzalez},
  {Guest}, {Hatch}, {Hess}, {Hoskin}, {Irwin}, {Knapen}, {Koposov}, {Kuchner},
  {Laigle}, {Lewis}, {Longhetti}, {Lucatello}, {M{\'e}ndez-Abreu}, {Mercurio},
  {Molaeinezhad}, {Mongui{\'o}}, {Morrison}, {Murphy}, {Peralta de Arriba},
  {P{\'e}rez}, {P{\'e}rez-R{\`a}fols}, {Pic{\'o}}, {Raddi}, {Romero-G{\'o}mez},
  {Royer}, {Siebert}, {Seabroke}, {Som}, {Terrett}, {Thomas}, {Wesson},
  {Worley}, {Alfaro}, {Prieto}, {Alonso-Santiago}, {Amos}, {Ashley},
  {Balaguer-N{\'u} nez}, {Balbinot}, {Bellazzini}, {Benn}, {Berlanas},
  {Bernard}, {Best}, {Bettoni}, {Bianco}, {Bishop}, {Blomqvist}, {Boeche},
  {Bolzonella}, {Bonoli}, {Bosma}, {Britavskiy}, {Busarello}, {Caffau},
  {Cantat-Gaudin}, {Castro-Ginard}, {Couto}, {Carbajo-Hijarrubia}, {Carter},
  {Casamiquela}, {Conrado}, {Corcho-Caballero}, {Costantin}, {Deason}, {de
  Burgos}, {De Grandi}, {Di Matteo}, {Dom{\'\i}nguez-G{\'o}mez}, {Dorda},
  {Drake}, {Dutta}, {Erkal}, {Feltzing}, {Ferr{\'e}-Mateu}, {Feuillet},
  {Figueras}, {Fossati}, {Franciosini}, {Frasca}, {Fumagalli}, {Gallazzi},
  {Garc{\'\i}a-Benito}, {Fusillo}, {Gebran}, {Gilbert}, {Gledhill},
  {Gonz{\'a}lez Delgado}, {Greimel}, {Guarcello}, {Guerra}, {Gullieuszik},
  {Haines}, {Hardcastle}, {Harris}, {Haywood}, {Helmi}, {Hernandez}, {Herrero},
  {Hughes}, {Irsic}, {Jablonka}, {Jarvis}, {Jordi}, {Kondapally}, {Kordopatis},
  {Krogager}, {La Barbera}, {Lam}, {Larsen}, {Lemasle}, {Lewis}, {Lhom{\'e}},
  {Lind}, {Lodi}, {Longobardi}, {Lonoce}, {Magrini}, {Ma{\'\i}z Apell{\'a}niz},
  {Marchal}, {Marco}, {Martin}, {Matsuno}, {Maurogordato}, {Merluzzi},
  {Miralda-Escud{\'e}}, {Molinari}, {Monari}, {Morelli}, {Mottram}, {Naylor},
  {Negueruela}, {Onorbe}, {Pancino}, {Peirani}, {Peletier}, {Pozzetti},
  {Rainer}, {Ramos}, {Read}, {Rossi}, {R{\"o}ttgering},
  {Rubi{\~n}o-Mart{\'\i}n}, {Sabater Montes}, {San Juan}, {Sanna}, {Schallig},
  {Schiavon}, {Schultheis}, {Serra}, {Shimwell}, {Sim{\'o}n-D{\'\i}az},
  {Smith}, {Sordo}, {Sorini}, {Soubiran}, {Starkenburg}, {Steele}, {Stott},
  {Stuik}, {Tolstoy}, {Tortora}, {Tsantaki}, {Van der Swaelmen}, {van Weeren},
  {Vergani}, {Verheijen}, {Verro}, {Vink}, {Vioque}, {Walcher}, {Walton},
  {Wegg}, {Weijmans}, {Williams}, {Wilson}, {Wright}, {Xylakis-Dornbusch},
  {Youakim}, {Zibetti}, \& {Zurita}}]{WEAVE}
{Jin}, S., {Trager}, S.~C., {Dalton}, G.~B., {et~al.} 2023, \mnras, 667, A103

\bibitem[{{Johnson}(1953)}]{Johnson53}
{Johnson}, H.~L. 1953, \apj, 117, 353

\bibitem[{{Kharchenko} \& {Roeser}(2009)}]{Kharchenko09}
{Kharchenko}, N.~V. \& {Roeser}, S. 2009, VizieR Online Data Catalog, I/280B

\bibitem[{{Kurucz}(1993{\natexlab{a}})}]{Kurucz1993a}
{Kurucz}, R. 1993{\natexlab{a}}, ATLAS9 Stellar Atmosphere Programs and 2 km/s
  grid. Kurucz CD-ROM No. 13. Cambridge, 13

\bibitem[{{Kurucz}(1993{\natexlab{b}})}]{Kurucz1993b}
{Kurucz}, R.~L. 1993{\natexlab{b}}, in Astronomical Society of the Pacific
  Conference Series, Vol.~44, IAU Colloq. 138: Peculiar versus Normal Phenomena
  in A-type and Related Stars, ed. M.~M. {Dworetsky}, F.~{Castelli}, \&
  R.~{Faraggiana}, 87

\bibitem[{{Kurucz} \& {Avrett}(1981)}]{Kurucz1981}
{Kurucz}, R.~L. \& {Avrett}, E.~H. 1981, SAO Special Report, 391

\bibitem[{{Lindegren} {et~al.}(2021){Lindegren}, {Klioner}, {Hern{\'a}ndez},
  {Bombrun}, {Ramos-Lerate}, {Steidelm{\"u}ller}, {Bastian}, {Biermann}, {de
  Torres}, {Gerlach}, {Geyer}, {Hilger}, {Hobbs}, {Lammers}, {McMillan},
  {Stephenson}, {Casta{\~n}eda}, {Davidson}, {Fabricius}, {Gracia-Abril},
  {Portell}, {Rowell}, {Teyssier}, {Torra}, {Bartolom{\'e}}, {Clotet},
  {Garralda}, {Gonz{\'a}lez-Vidal}, {Torra}, {Abbas}, {Altmann}, {Anglada
  Varela}, {Balaguer-N{\'u}{\~n}ez}, {Balog}, {Barache}, {Becciani}, {Bernet},
  {Bertone}, {Bianchi}, {Bouquillon}, {Brown}, {Bucciarelli}, {Busonero},
  {Butkevich}, {Buzzi}, {Cancelliere}, {Carlucci}, {Charlot}, {Cioni},
  {Crosta}, {Crowley}, {del Peloso}, {del Pozo}, {Drimmel}, {Esquej}, {Fienga},
  {Fraile}, {Gai}, {Garcia-Reinaldos}, {Guerra}, {Hambly}, {Hauser},
  {Jan{\ss}en}, {Jordan}, {Kostrzewa-Rutkowska}, {Lattanzi}, {Liao}, {Licata},
  {Lister}, {L{\"o}ffler}, {Marchant}, {Masip}, {Mignard}, {Mints}, {Molina},
  {Mora}, {Morbidelli}, {Murphy}, {Pagani}, {Panuzzo}, {Pe{\~n}alosa Esteller},
  {Poggio}, {Re Fiorentin}, {Riva}, {Sagrist{\`a} Sell{\'e}s}, {Sanchez
  Gimenez}, {Sarasso}, {Sciacca}, {Siddiqui}, {Smart}, {Souami}, {Spagna},
  {Steele}, {Taris}, {Utrilla}, {van Reeven}, \& {Vecchiato}}]{Lindegren21}
{Lindegren}, L., {Klioner}, S.~A., {Hern{\'a}ndez}, J., {et~al.} 2021, \aap,
  649, A2

\bibitem[{{Magrini} {et~al.}(2023){Magrini}, {Viscasillas V{\'a}zquez},
  {Spina}, {Randich}, {Romano}, {Franciosini}, {Recio-Blanco}, {Nordlander},
  {D'Orazi}, {Baratella}, {Smiljanic}, {Dantas}, {Pasquini}, {Spitoni},
  {Casali}, {Van der Swaelmen}, {Bensby}, {Stonkute}, {Feltzing}, {Sacco},
  {Bragaglia}, {Pancino}, {Heiter}, {Biazzo}, {Gilmore}, {Bergemann},
  {Tautvai{\v{s}}ien{\.{e}}}, {Worley}, {Hourihane}, {Gonneau}, \&
  {Morbidelli}}]{Magrini23}
{Magrini}, L., {Viscasillas V{\'a}zquez}, C., {Spina}, L., {et~al.} 2023, \aap,
  669, A119

\bibitem[{{Majewski} {et~al.}(2017){Majewski}, {Schiavon}, {Frinchaboy},
  {Allende Prieto}, {Barkhouser}, {Bizyaev}, {Blank}, {Brunner}, {Burton},
  {Carrera}, {Chojnowski}, {Cunha}, {Epstein}, {Fitzgerald}, {Garc{\'\i}a
  P{\'e}rez}, {Hearty}, {Henderson}, {Holtzman}, {Johnson}, {Lam}, {Lawler},
  {Maseman}, {M{\'e}sz{\'a}ros}, {Nelson}, {Nguyen}, {Nidever}, {Pinsonneault},
  {Shetrone}, {Smee}, {Smith}, {Stolberg}, {Skrutskie}, {Walker}, {Wilson},
  {Zasowski}, {Anders}, {Basu}, {Beland}, {Blanton}, {Bovy}, {Brownstein},
  {Carlberg}, {Chaplin}, {Chiappini}, {Eisenstein}, {Elsworth}, {Feuillet},
  {Fleming}, {Galbraith-Frew}, {Garc{\'\i}a}, {Garc{\'\i}a-Hern{\'a}ndez},
  {Gillespie}, {Girardi}, {Gunn}, {Hasselquist}, {Hayden}, {Hekker}, {Ivans},
  {Kinemuchi}, {Klaene}, {Mahadevan}, {Mathur}, {Mosser}, {Muna}, {Munn},
  {Nichol}, {O'Connell}, {Parejko}, {Robin}, {Rocha-Pinto}, {Schultheis},
  {Serenelli}, {Shane}, {Silva Aguirre}, {Sobeck}, {Thompson}, {Troup},
  {Weinberg}, \& {Zamora}}]{APOGEE}
{Majewski}, S.~R., {Schiavon}, R.~P., {Frinchaboy}, P.~M., {et~al.} 2017, \aj,
  154, 94

\bibitem[{{Mamajek} \& {Hillenbrand}(2008)}]{Mamajek2008}
{Mamajek}, E.~E. \& {Hillenbrand}, L.~A. 2008, \apj, 687, 1264

\bibitem[{{Manteiga} {et~al.}(1991){Manteiga}, {Martinez-Roger}, {Morales}, \&
  {Sabau}}]{Manteiga91}
{Manteiga}, M., {Martinez-Roger}, C., {Morales}, C., \& {Sabau}, L. 1991,
  \aaps, 87, 419

\bibitem[{{Marigo} {et~al.}(2017){Marigo}, {Girardi}, {Bressan}, {Rosenfield},
  {Aringer}, {Chen}, {Dussin}, {Nanni}, {Pastorelli}, {Rodrigues}, {Trabucchi},
  {Bladh}, {Dalcanton}, {Groenewegen}, {Montalb{\'a}n}, \& {Wood}}]{Marigo17}
{Marigo}, P., {Girardi}, L., {Bressan}, A., {et~al.} 2017, \apj, 835, 77

\bibitem[{{McNamara} \& {Sanders}(1977)}]{McNamara77}
{McNamara}, B.~J. \& {Sanders}, W.~L. 1977, \aaps, 30, 45

\bibitem[{{Mermilliod} {et~al.}(2009){Mermilliod}, {Mayor}, \&
  {Udry}}]{Mermilliod09}
{Mermilliod}, J.~C., {Mayor}, M., \& {Udry}, S. 2009, \aap, 498, 949

\bibitem[{{Mohan} \& {Sagar}(1985)}]{Mohan85}
{Mohan}, V. \& {Sagar}, R. 1985, \mnras, 213, 337

\bibitem[{{Myers} {et~al.}(2022){Myers}, {Donor}, {Spoo}, {Frinchaboy},
  {Cunha}, {Price-Whelan}, {Majewski}, {Beaton}, {Zasowski}, {O'Connell},
  {Ray}, {Bizyaev}, {Chiappini}, {Garc{\'\i}a-Hern{\'a}ndez}, {Geisler},
  {J{\"o}nsson}, {Lane}, {Longa-Pe{\~n}a}, {Minchev}, {Minniti}, {Nitschelm},
  \& {Roman-Lopes}}]{Myers22}
{Myers}, N., {Donor}, J., {Spoo}, T., {et~al.} 2022, \aj, 164, 85

\bibitem[{{Platais}(1994)}]{Platais94}
{Platais}, I. 1994, Bulletin d'Information du Centre de Donnees Stellaires, 44,
  9

\bibitem[{{Randich} {et~al.}(2022){Randich}, {Gilmore}, {Magrini}, {Sacco},
  {Jackson}, {Jeffries}, {Worley}, {Hourihane}, {Gonneau}, {Viscasillas
  Vazquez}, {Franciosini}, {Lewis}, {Alfaro}, {Allende Prieto}, {Bensby},
  {Blomme}, {Bragaglia}, {Flaccomio}, {Fran{\c{c}}ois}, {Irwin}, {Koposov},
  {Korn}, {Lanzafame}, {Pancino}, {Recio-Blanco}, {Smiljanic}, {Van Eck},
  {Zwitter}, {Asplund}, {Bonifacio}, {Feltzing}, {Binney}, {Drew}, {Ferguson},
  {Micela}, {Negueruela}, {Prusti}, {Rix}, {Vallenari}, {Bayo}, {Bergemann},
  {Biazzo}, {Carraro}, {Casey}, {Damiani}, {Frasca}, {Heiter}, {Hill},
  {Jofr{\'e}}, {de Laverny}, {Lind}, {Marconi}, {Martayan}, {Masseron},
  {Monaco}, {Morbidelli}, {Prisinzano}, {Sbordone}, {Sousa}, {Zaggia},
  {Adibekyan}, {Bonito}, {Caffau}, {Daflon}, {Feuillet}, {Gebran}, {Gonzalez
  Hernandez}, {Guiglion}, {Herrero}, {Lobel}, {Maiz Apellaniz}, {Merle},
  {Mikolaitis}, {Montes}, {Morel}, {Soubiran}, {Spina}, {Tabernero},
  {Tautvai{\v{s}}iene}, {Traven}, {Valentini}, {Van der Swaelmen}, {Villanova},
  {Wright}, {Abbas}, {Aguirre B{\o}rsen-Koch}, {Alves}, {Balaguer-Nunez},
  {Barklem}, {Barrado}, {Berlanas}, {Binks}, {Bressan}, {Capuzzo-Dolcetta},
  {Casagrande}, {Casamiquela}, {Collins}, {D'Orazi}, {Dantas}, {Debattista},
  {Delgado-Mena}, {Di Marcantonio}, {Drazdauskas}, {Evans}, {Famaey},
  {Franchini}, {Fr{\'e}mat}, {Friel}, {Fu}, {Geisler}, {Gerhard}, {Gonzalez
  Solares}, {Grebel}, {Gutierrez Albarran}, {Hatzidimitriou}, {Held},
  {Jim{\'e}nez-Esteban}, {J{\"o}nsson}, {Jordi}, {Khachaturyants},
  {Kordopatis}, {Kos}, {Lagarde}, {Mahy}, {Mapelli}, {Marfil}, {Martell},
  {Messina}, {Miglio}, {Minchev}, {Moitinho}, {Montalban}, {Monteiro},
  {Morossi}, {Mowlavi}, {Mucciarelli}, {Murphy}, {Nardetto}, {Ortolani},
  {Paletou}, {Palou{\v{s}}}, {Paunzen}, {Pickering}, {Quirrenbach}, {Re
  Fiorentin}, {Read}, {Romano}, {Ryde}, {Sanna}, {Santos}, {Seabroke},
  {Spagna}, {Steinmetz}, {Stonkut{\'e}}, {Sutorius}, {Th{\'e}venin}, {Tosi},
  {Tsantaki}, {Vink}, {Wright}, {Wyse}, {Zoccali}, {Zorec}, {Zucker}, \&
  {Walton}}]{Randich22}
{Randich}, S., {Gilmore}, G., {Magrini}, L., {et~al.} 2022, \aap, 666, A121

\bibitem[{{Sestito} \& {Randich}(2005)}]{Sestito2005}
{Sestito}, P. \& {Randich}, S. 2005, \aap, 442, 615

\bibitem[{{Skrutskie} {et~al.}(2006){Skrutskie}, {Cutri}, {Stiening},
  {Weinberg}, {Schneider}, {Carpenter}, {Beichman}, {Capps}, {Chester},
  {Elias}, {Huchra}, {Liebert}, {Lonsdale}, {Monet}, {Price}, {Seitzer},
  {Jarrett}, {Kirkpatrick}, {Gizis}, {Howard}, {Evans}, {Fowler}, {Fullmer},
  {Hurt}, {Light}, {Kopan}, {Marsh}, {McCallon}, {Tam}, {Van Dyk}, \&
  {Wheelock}}]{2MASS}
{Skrutskie}, M.~F., {Cutri}, R.~M., {Stiening}, R., {et~al.} 2006, \aj, 131,
  1163

\bibitem[{{Smiljanic} {et~al.}(2014){Smiljanic}, {Korn}, {Bergemann}, {Frasca},
  {Magrini}, {Masseron}, {Pancino}, {Ruchti}, {San Roman}, {Sbordone}, {Sousa},
  {Tabernero}, {Tautvai{\v{s}}ien{\.{e}}}, {Valentini}, {Weber}, {Worley},
  {Adibekyan}, {Allende Prieto}, {Barisevi{\v{c}}ius}, {Biazzo},
  {Blanco-Cuaresma}, {Bonifacio}, {Bragaglia}, {Caffau}, {Cantat-Gaudin},
  {Chorniy}, {de Laverny}, {Delgado-Mena}, {Donati}, {Duffau}, {Franciosini},
  {Friel}, {Geisler}, {Gonz{\'a}lez Hern{\'a}ndez}, {Gruyters}, {Guiglion},
  {Hansen}, {Heiter}, {Hill}, {Jacobson}, {Jofre}, {J{\"o}nsson}, {Lanzafame},
  {Lardo}, {Ludwig}, {Maiorca}, {Mikolaitis}, {Montes}, {Morel}, {Mucciarelli},
  {Mu{\~n}oz}, {Nordlander}, {Pasquini}, {Puzeras}, {Recio-Blanco}, {Ryde},
  {Sacco}, {Santos}, {Serenelli}, {Sordo}, {Soubiran}, {Spina}, {Steffen},
  {Vallenari}, {Van Eck}, {Villanova}, {Gilmore}, {Randich}, {Asplund},
  {Binney}, {Drew}, {Feltzing}, {Ferguson}, {Jeffries}, {Micela}, {Negueruela},
  {Prusti}, {Rix}, {Alfaro}, {Babusiaux}, {Bensby}, {Blomme}, {Flaccomio},
  {Fran{\c{c}}ois}, {Irwin}, {Koposov}, {Walton}, {Bayo}, {Carraro}, {Costado},
  {Damiani}, {Edvardsson}, {Hourihane}, {Jackson}, {Lewis}, {Lind}, {Marconi},
  {Martayan}, {Monaco}, {Morbidelli}, {Prisinzano}, \& {Zaggia}}]{Smiljanic14}
{Smiljanic}, R., {Korn}, A.~J., {Bergemann}, M., {et~al.} 2014, \aap, 570, A122

\bibitem[{{Soderblom} {et~al.}(1993){Soderblom}, {Jones}, {Balachandran},
  {Stauffer}, {Duncan}, {Fedele}, \& {Hudon}}]{Soderblom93}
{Soderblom}, D.~R., {Jones}, B.~F., {Balachandran}, S., {et~al.} 1993, \aj,
  106, 1059

\bibitem[{{Telting} {et~al.}(2014){Telting}, {Avila}, {Buchhave}, {Frandsen},
  {Gandolfi}, {Lindberg}, {Stempels}, {Prins}, \& {NOT staff}}]{Telting14}
{Telting}, J.~H., {Avila}, G., {Buchhave}, L., {et~al.} 2014, Astronomische
  Nachrichten, 335, 41

\bibitem[{{Trumpler}(1928)}]{Trumpler28}
{Trumpler}, R.~J. 1928, \pasp, 40, 265

\bibitem[{{Weaver}(1953)}]{Weaver53}
{Weaver}, H.~F. 1953, \apj, 117, 366

\bibitem[{{Zhang} {et~al.}(2022){Zhang}, {Lucatello}, {Bragaglia},
  {Alonso-Santiago}, {Andreuzzi}, {Casali}, {Carrera}, {Carretta}, {D'Orazi},
  {Frasca}, {Fu}, {Magrini}, {Minchev}, {Origlia}, {Spina}, \&
  {Vallenari}}]{Zhang22}
{Zhang}, R., {Lucatello}, S., {Bragaglia}, A., {et~al.} 2022, \aap, 667, A103

\bibitem[{{Zhang} {et~al.}(2021){Zhang}, {Lucatello}, {Bragaglia}, {Carrera},
  {Spina}, {Alonso-Santiago}, {Andreuzzi}, {Casali}, {Carretta}, {Frasca},
  {Fu}, {Magrini}, {Origlia}, {D'Orazi}, \& {Vallenari}}]{Zhang21}
{Zhang}, R., {Lucatello}, S., {Bragaglia}, A., {et~al.} 2021, \aap, 654, A77

\end{thebibliography}



\appendix

\section{Additional materials}\label{appendix}


\begin{table*}
\caption{$Gaia$ DR3 astrometric data and cluster membership, according to \citet[][P$_{Pl}$]{Platais94} and \citet[][P$_{CG}$]{CG18}, for all stars observed spectroscopically in this work (P$_{TW}$).}
\label{tab_astrom}
\begin{center}
\begin{tabular}{lcccccccccc}   
\hline\hline
\multirow{2}{*}{Star} & \multirow{2}{*}{$Gaia$-DR3 ID} &  RA     &   DEC   &      r       & \multirow{2}{*}{P$_{Pl}$} & \multirow{2}{*}{P$_{CG}$} & \multirow{2}{*}{P$_{TW}$}    & $\varpi$ &  $\mu_{\alpha*}$  & $\mu_{\delta}$    \\ 
                      &                                & (J2000) & (J2000) &  ($\arcmin$) &                           &                           &                              &  (mas)   & (mas\,a$^{-1}$)  &  (mas\,a$^{-1}$) \\ 
\hline
\multicolumn{11}{c}{Members}\\    
\hline
0305  &  \scriptsize{1978587946054581760}  &  321.881658501  &  48.322025012  &  40.5  &  0.72  &  1.0  &  I    &  3.397  &  $-$7.488  &  $-$20.357  \\ 
2451  &  \scriptsize{1978555647899068544}  &  322.578899973  &  48.390844737  &  15.1  &  0.97  &  0.9  &  I    &  3.367  &  $-$7.856  &  $-$19.652  \\ 
3061  &  \scriptsize{1978556644336044416}  &  322.725446694  &  48.427086634  &  12.6  &  0.70  &  1.0  &  I    &  3.328  &  $-$7.537  &  $-$19.258  \\ 
3288  &  \scriptsize{1978699683923650688}  &  322.782240751  &  48.644378692  &  24.2  &  0.77  &  1.0  &  I    &  3.432  &  $-$7.362  &  $-$20.054  \\ 
3311  &  \scriptsize{1978533142270627328}  &  322.793918577  &  48.070524115  &  11.3  &  0.98  &  1.0  &  I    &  3.415  &  $-$7.497  &  $-$19.837  \\ 
3781  &  \scriptsize{1978652748518359808}  &  322.926601311  &  48.584617826  &  20.3  &  0.97  &  1.0  &  I    &  3.324  &  $-$7.769  &  $-$19.771  \\ 
3814  &  \scriptsize{1978647762043764608}  &  322.936042739  &  48.484353871  &  14.4  &  0.98  &  0.7  &  I    &  3.308  &  $-$8.004  &  $-$19.684  \\ 
4265  &  \scriptsize{1978656218831726848}  &  323.061071758  &  48.639501413  &  24.5  &  0.97  &  1.0  &  I    &  3.291  &  $-$7.428  &  $-$19.568  \\ 
4322  &  \scriptsize{1978628593621993088}  &  323.081133228  &  48.200479032  &   8.2  &  0.75  &  1.0  &  I    &  3.332  &  $-$7.743  &  $-$20.166  \\ 
4438  &  \scriptsize{1978740434565656320}  &  323.114228026  &  49.181066745  &  56.7  &  0.78  &  1.0  &  I    &  3.340  &  $-$7.476  &  $-$19.720  \\ 
4673  &  \scriptsize{1978641856481498752}  &  323.179813508  &  48.483167010  &  18.3  &  0.98  &  1.0  &  I    &  3.380  &  $-$7.614  &  $-$19.973  \\ 
5045  &  \scriptsize{1978443325910675712}  &  323.292127855  &  48.303406860  &  16.5  &  0.98  & \dots &  I    &  3.325  &  $-$8.080  &  $-$19.576  \\ 
5609  &  \scriptsize{1978742908476235776}  &  323.448921752  &  49.160993150  &  59.2  &  0.42  & \dots &  I    &  3.154  &  $-$7.165  &  $-$19.645  \\ 
5729  &  \scriptsize{1978743355145551104}  &  323.486420337  &  49.196906735  &  61.7  &  0.70  &  1.0  &  I    &  3.308  &  $-$7.333  &  $-$19.552  \\ 
6791  &  \scriptsize{1978460471421230848}  &  323.805577079  &  48.437269774  &  38.3  &  0.97  &  1.0  &  I    &  3.413  &  $-$7.613  &  $-$20.013  \\ 
7140  &  \scriptsize{1978404499404948096}  &  323.948575257  &  48.108247352  &  43.2  &  0.96  &  1.0  &  I    &  3.363  &  $-$7.719  &  $-$19.540  \\ 
ucac  &  \scriptsize{1972328941741362944}  &  320.502548884  &  46.489037684  & 143.3  &  \dots &  1.0  &  I    &  3.234  &  $-$7.011  &  $-$19.731  \\ 
\hline
\multicolumn{11}{c}{Non-members}\\ 
\hline
3004  &  \scriptsize{1978557640768446464}  &  322.710204334  &  48.483127406  &  15.9  &  0.95  & \dots & \dots &  3.112  &  $-$7.358  &  $-$16.743  \\ 
3423  &  \scriptsize{1978652508006151296}  &  322.825494272  &  48.623094206  &  22.7  &  0.72  & \dots & \dots &  3.050  &  $-$6.599  &  $-$22.305  \\ 
4294  &  \scriptsize{1978636290203508608}  &  323.071129883  &  48.443836401  &  13.9  &  0.96  & \dots & \dots &  3.056  &  $-$5.833  &  $-$22.329  \\ 
\hline

\hline
 
\end{tabular}
\end{center}
\end{table*}

\begin{table*}
\caption{Photometry for stars observed spectroscopically in this work.}
\label{tab_fotom}
\begin{center}
\begin{tabular}{lccccccccc}   
\hline\hline
Star  & $V^a$ & $(B-V)^a$ & $J^b$   &  $H^b$  & $K_{\textrm{S}}^b$ & $G^c$ & $(G_{\textrm{BP}}-G_{\textrm{RP}})^c$ & $E(G_{\textrm{BP}}-G_{\textrm{RP}})^c$ &  $A_G^c$     \\  
\hline
\multicolumn{10}{c}{Members}\\ 
\hline
0305  &  12.359  &  0.654  &  11.178  &  10.854  &  10.813  &  12.231  &     0.836  &  0.000  &  0.002  \\
2451  &   7.318$^*$  &  0.121$^*$  &   7.312  &   7.353  &   7.351  &   7.358  &  $-$0.023  &  0.041  &  0.076  \\
3061  &  11.256  &  0.466  &  10.329  &  10.141  &  10.097  &  11.141  &     0.620  &  0.001  &  0.001  \\
3288  &  12.698  &  0.751  &  11.474  &  11.144  &  11.075  &  12.602  &     0.889  &  0.023  &  0.042  \\
3311  &  10.168  &  0.296  &   9.587  &   9.524  &   9.468  &  10.110  &     0.396  &  0.032  &  0.060  \\
3781  &   6.961$^*$  &  0.139$^*$  &   6.826  &   6.846  &   6.890  &   6.831  &  $-$0.036  &  0.051  &  0.096  \\
3814  &   7.630$^*$  &  0.057$^*$  &   7.663  &   7.608  &   7.668  &   7.663  &  $-$0.004  &  0.040  &  0.075  \\
4265  &   8.228  &  0.029  &   8.104  &   8.084  &   8.107  &   8.252  &     0.056  &  0.132  &  0.247  \\
4322  &  12.403  &  0.646  &  11.193  &  10.866  &  10.826  &  12.249  &     0.832  &  0.001  &  0.001  \\
4438  &  11.447  &  0.535  &  10.503  &  10.288  &  10.223  &  11.329  &     0.651  &  0.003  &  0.005  \\
4673  &   9.008  &  0.082  &   8.874  &   8.898  &   8.875  &   9.035  &     0.088  &  0.110  &  0.206  \\
5045  &   6.691$^*$  &  0.195$^*$  &   6.697  &   6.778  &   6.728  &   6.840  &     0.012  &  0.048  &  0.091  \\
5609  &  10.460  &  0.308  &   9.713  &   9.567  &   9.521  &  10.343  &     0.477  &  0.016  &  0.029  \\
5729  &   9.356  &  0.176  &   9.041  &   8.997  &   8.993  &   9.369  &     0.204  &  0.032  &  0.060  \\
6791  &   9.565  &  0.286  &   9.267  &   9.142  &   9.181  &   9.562  &     0.220  &  \dots  &  \dots  \\
7140  &   9.392  &  0.355  &   9.150  &   9.132  &   9.072  &   9.405  &     0.179  &  0.022  &  0.041  \\
ucac  &  12.649  &  0.633  &  11.403  &  11.098  &  11.010  &  12.480  &     0.847  &  0.058  &  0.108  \\
\hline
\multicolumn{10}{c}{Non-members}\\ 
\hline
3004  &  10.428  &  0.354  &   9.927  &   9.793  &   9.746  &  10.526  &     0.466  &  0.024  &  0.045  \\
3423  &  10.932  &  0.508  &  10.013  &   9.673  &   9.674  &  10.813  &     0.668  &  0.001  &  0.003  \\
4294  &   6.813  &  0.207  &   6.478  &   6.470  &   6.513  &   6.573  &     0.018  &  \dots  &  \dots  \\
\hline
\end{tabular}
\end{center}
{\bf Notes.} $^a$ From the APASS catalogue \citep{APASS}. $^b$ From the 2MASS catalogue \citep{2MASS}. $^c$ From the $Gaia$-DR3 catalogue \citep{DR3}. $^*$ Scaled from the $V$ and $(B-V)$ values of \citet{Kharchenko09} and the relation proposed in the present work.
\end{table*}


\begin{figure*}[ht]
\begin{center}

\includegraphics[width=15cm]{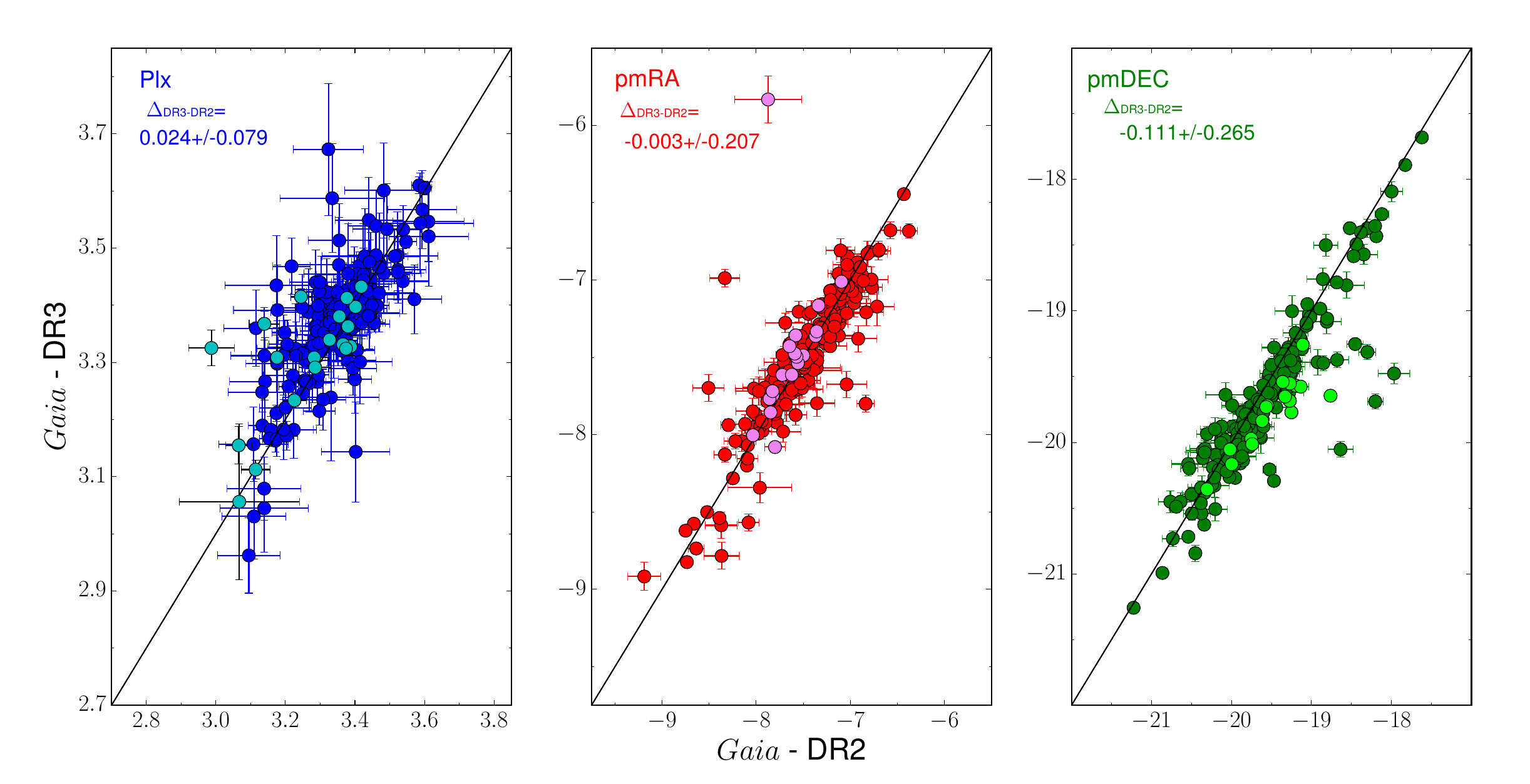}	
\caption{Astrometric data for the cluster members identified by \citet{CG18} in $Gaia$-DR2 and $Gaia$-DR3. The lightest colour in each panel represents the 
stars observed in this work. For each parameter the mean difference (DR3-DR2) and its dispersion are indicated in the upper left part of the plots.
}
\label{astrom_dr2dr3}
\end{center}
\end{figure*}

\begin{figure*}
\begin{center}

\includegraphics[width=11cm]{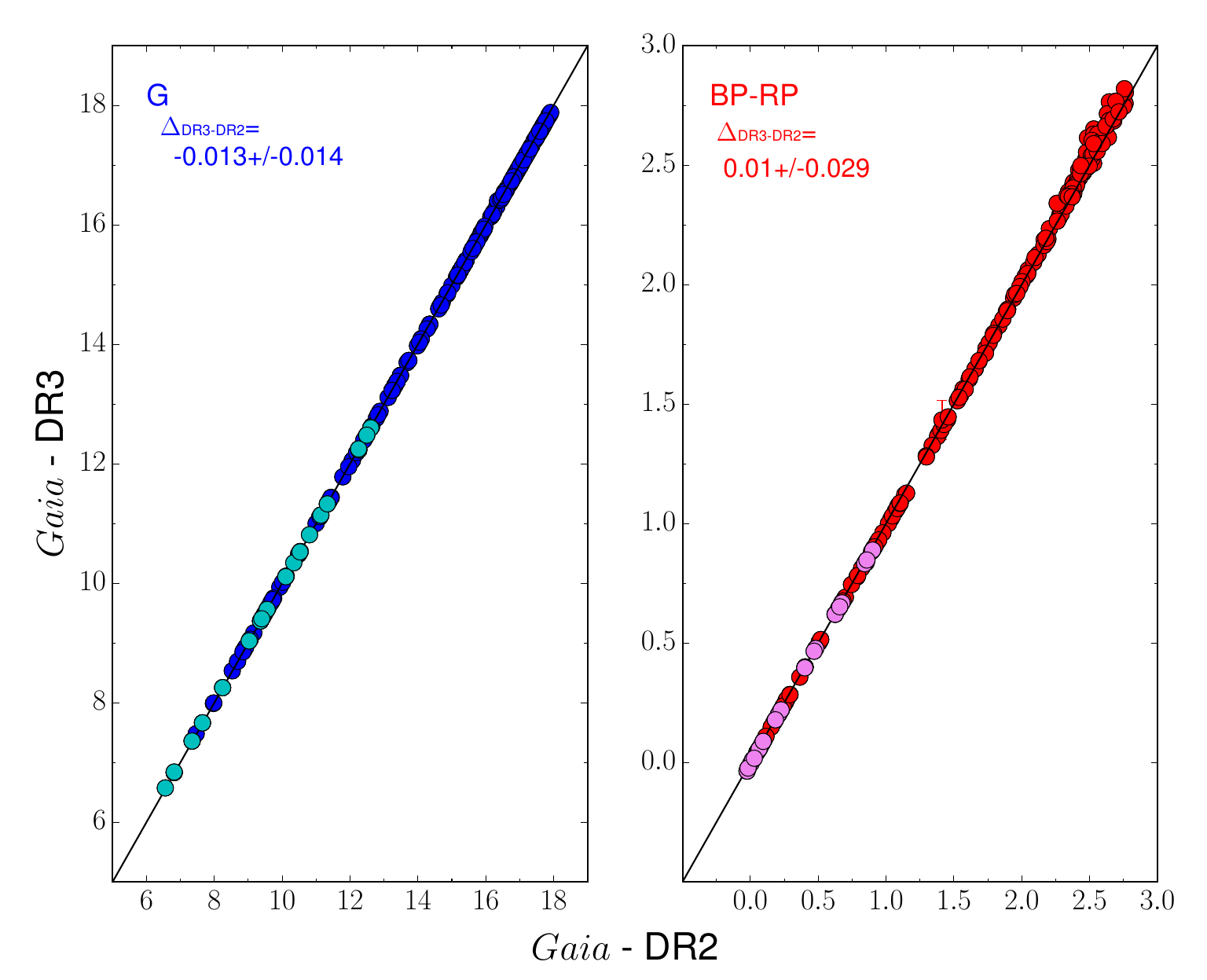}	
\caption{The same as in Fig.~\ref{astrom_dr2dr3} but with the photometric values.}

\label{photom_dr2dr3}
\end{center}
\end{figure*}

\begin{figure*}
\begin{center}

\includegraphics[width=6cm]{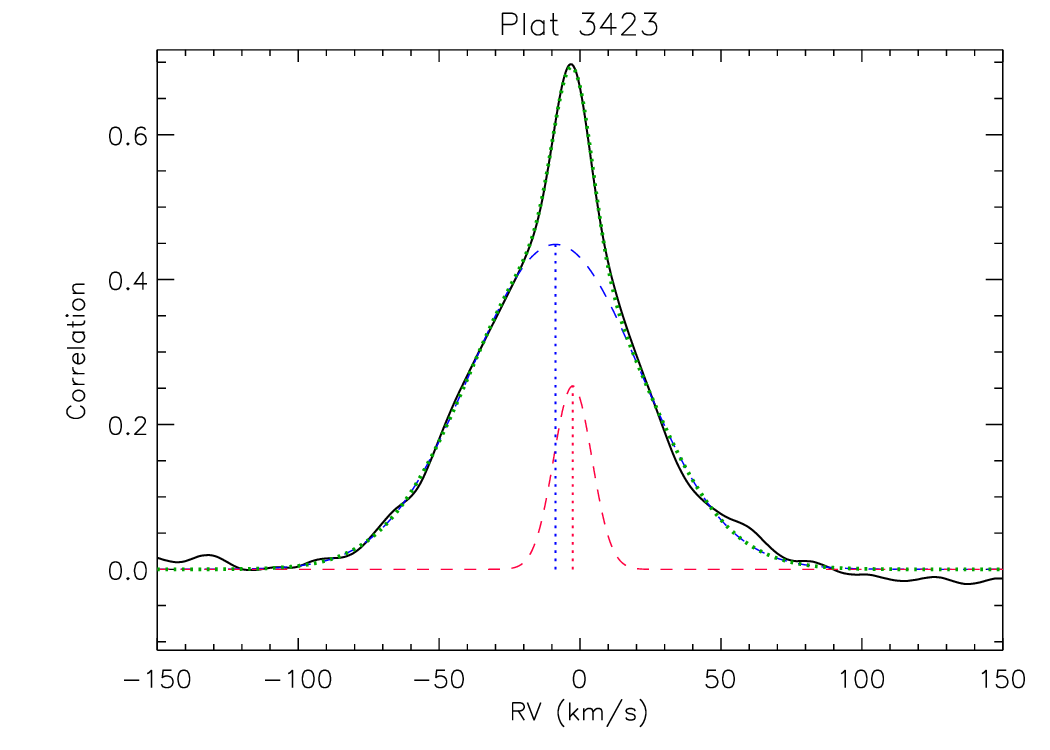}	
\includegraphics[width=6cm]{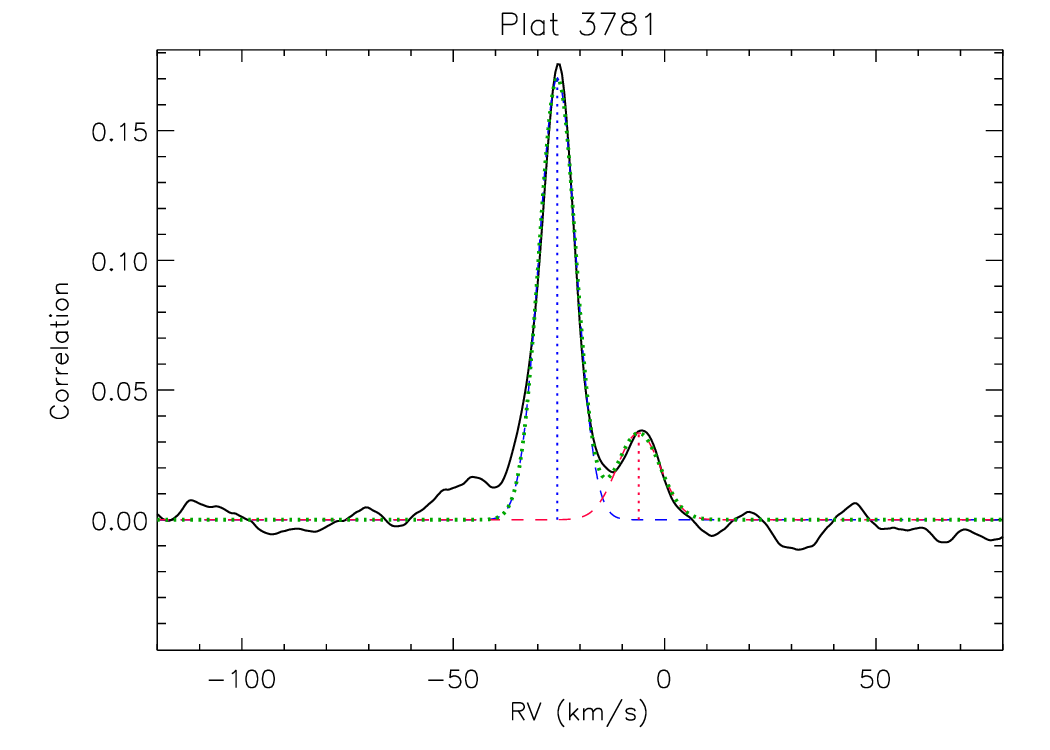}	
\includegraphics[width=6cm]{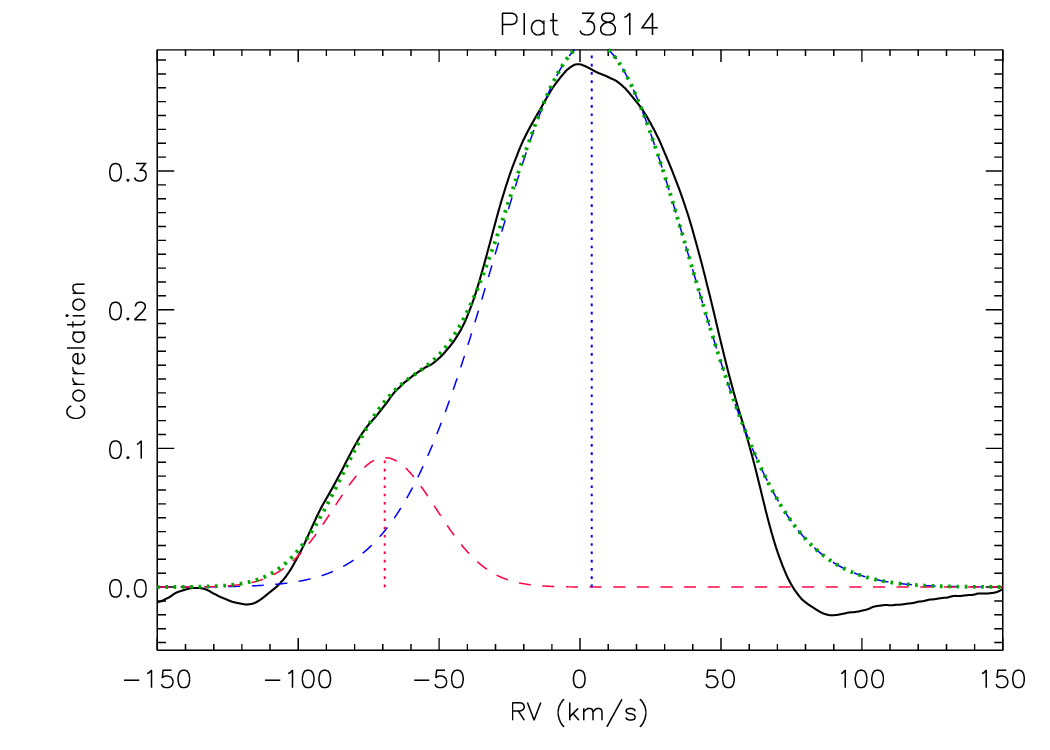}	
\caption{Cross-correlation functions (CCFs) for the three newly discovered spectroscopic binaries. In each box, the CCF is depicted with a full black line. 
The two-Gaussian fit is over-plotted with a dotted green line, while the Gaussians reproducing the primary (more luminous) and secondary components are plotted with dashed blue and red lines, respectively. 
}
\label{Fig:CCF_Plat3423}
\end{center}
\end{figure*}

\begin{figure*}
\begin{center}

\includegraphics[width=11cm]{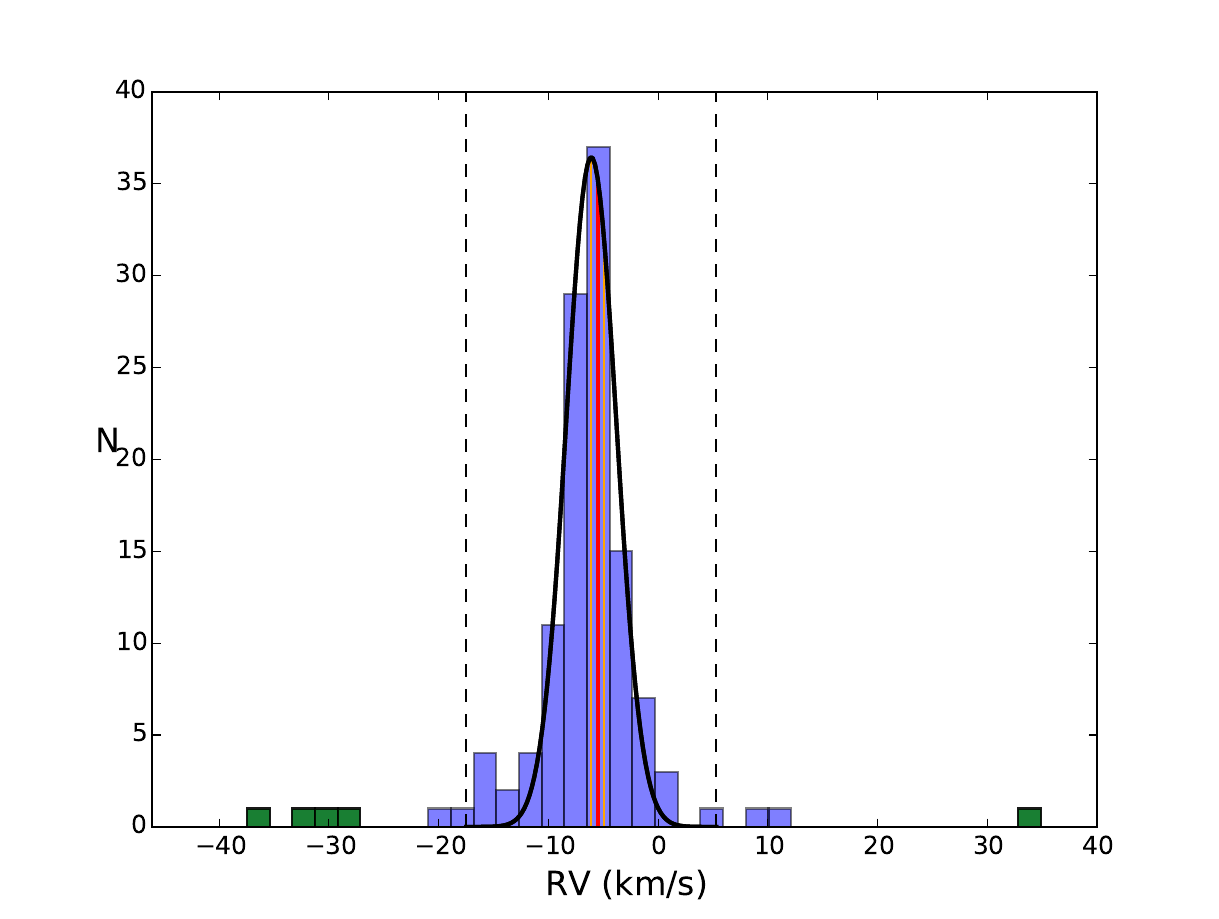}	
\caption{Distribution of the $Gaia$ RV for the likely candidate members selected on the basis of the astrometric and photometric criteria explained in the text (blue histogram). The average RV for the cluster, found by the Gaussian fitting (solid black line), peaks at $-$6.1$\pm$2.3\,km\,s$^{-1}$. This value agrees closely with the RV obtained from the stars observed spectroscopically in this work (red line for the average value and orange lines for its uncertainty). 
The 5$\sigma$ boundary is marked by dashed lines. The five objects that lie outside this boundary, taking into account their errors, have been represented with a green histogram and are considered non-members.}
\label{hist_RV}
\end{center}
\end{figure*}

\begin{table*}
\caption{Chemical abundances, expressed as $A(X)$=log[$n(X)/n(H)$]+12, for the cool stars observed in this work.}
\label{tab_ind_abund}
\begin{center}
\begin{tabular}{lccccccccc}   
\hline\hline
X  &       0305      &     3004$^*$    &       3061       &      3288       &      3311       &      4322       &      4438       &      5609       &      ucac      \\  
\hline         
       
C  &  8.52$\pm$0.10  &  8.78$\pm$0.12  &  8.70$\pm$0.09  &  8.49$\pm$0.12  &  8.26$\pm$0.15  &  8.31$\pm$0.11  &  8.49$\pm$0.10  &  8.53$\pm$0.07  &  8.70$\pm$0.14  \\
Na &  5.93$\pm$0.10  &  6.02$\pm$0.10  &  5.70$\pm$0.10  &  6.04$\pm$0.09  &  5.73$\pm$0.14  &  5.90$\pm$0.10  &  6.00$\pm$0.13  &  6.06$\pm$0.12  &  5.99$\pm$0.15  \\
Mg &  7.56$\pm$0.10  &  7.61$\pm$0.10  &  7.63$\pm$0.10  &  7.69$\pm$0.10  &  7.63$\pm$0.15  &  7.62$\pm$0.10  &  7.72$\pm$0.15  &  7.77$\pm$0.20  &  7.67$\pm$0.10  \\
Al &  6.53$\pm$0.10  &  6.61$\pm$0.15  &  6.32$\pm$0.10  &  6.59$\pm$0.10  &  6.47$\pm$0.10  &  6.60$\pm$0.10  &  6.54$\pm$0.09  &  6.50$\pm$0.12  &  6.62$\pm$0.10  \\
Si &  7.39$\pm$0.12  &  7.85$\pm$0.11  &  7.39$\pm$0.08  &  7.53$\pm$0.10  &  7.62$\pm$0.10  &  7.68$\pm$0.11  &  7.63$\pm$0.09  &  7.59$\pm$0.10  &  7.54$\pm$0.12  \\
S  &  7.37$\pm$0.11  &  7.57$\pm$0.14  &  7.22$\pm$0.11  &  7.54$\pm$0.10  &  7.45$\pm$0.09  &  7.48$\pm$0.10  &  7.41$\pm$0.07  &  7.54$\pm$0.07  &  7.61$\pm$0.10  \\
Ca &  6.62$\pm$0.10  &  6.64$\pm$0.05  &  6.35$\pm$0.11  &  6.63$\pm$0.08  &  6.44$\pm$0.03  &  6.51$\pm$0.11  &  6.53$\pm$0.09  &  6.59$\pm$0.09  &  6.61$\pm$0.08  \\
Sc &  3.28$\pm$0.09  &  3.49$\pm$0.12  &  3.36$\pm$0.05  &  3.36$\pm$0.10  &  3.17$\pm$0.07  &  3.28$\pm$0.10  &  3.39$\pm$0.10  &  3.29$\pm$0.09  &  3.33$\pm$0.10  \\
Ti &  4.93$\pm$0.12  &  5.21$\pm$0.10  &  4.63$\pm$0.13  &  5.00$\pm$0.09  &  5.17$\pm$0.09  &  4.81$\pm$0.11  &  5.02$\pm$0.08  &  5.18$\pm$0.10  &  4.83$\pm$0.10  \\
V  &  4.13$\pm$0.08  &  4.30$\pm$0.10  &  4.15$\pm$0.17  &  4.23$\pm$0.08  &  4.27$\pm$0.04  &  4.17$\pm$0.05  &  4.22$\pm$0.10  &  4.23$\pm$0.10  &  4.24$\pm$0.09  \\
Cr &  5.79$\pm$0.04  &  5.89$\pm$0.10  &  5.57$\pm$0.13  &  5.86$\pm$0.10  &  5.80$\pm$0.06  &  5.78$\pm$0.04  &  5.79$\pm$0.11  &  5.75$\pm$0.08  &  5.84$\pm$0.10  \\
Mn &  5.48$\pm$0.10  &  5.65$\pm$0.10  &  5.43$\pm$0.10  &  5.59$\pm$0.10  &  5.35$\pm$0.05  &  5.49$\pm$0.12  &  5.58$\pm$0.06  &  5.62$\pm$0.10  &  5.59$\pm$0.10  \\
Fe &  7.48$\pm$0.13  &  7.53$\pm$0.14  &  7.05$\pm$0.10  &  7.58$\pm$0.12  &  7.51$\pm$0.11  &  7.48$\pm$0.12  &  7.52$\pm$0.17  &  7.33$\pm$0.12  &  7.51$\pm$0.10  \\
Co &  4.86$\pm$0.14  &  5.25$\pm$0.10  &  4.76$\pm$0.13  &  4.98$\pm$0.12  &  4.95$\pm$0.18  &  4.91$\pm$0.10  &  4.97$\pm$0.10  &  5.07$\pm$0.10  &  4.97$\pm$0.10  \\
Ni &  6.16$\pm$0.12  &  6.39$\pm$0.05  &  6.06$\pm$0.12  &  6.24$\pm$0.11  &  6.26$\pm$0.09  &  6.16$\pm$0.13  &  6.25$\pm$0.13  &  6.28$\pm$0.10  &  6.21$\pm$0.11  \\
Cu &  4.34$\pm$0.10  &  4.48$\pm$0.10  &  4.21$\pm$0.10  &  4.36$\pm$0.10  &  4.36$\pm$0.10  &  4.27$\pm$0.10  &  4.33$\pm$0.10  &  4.50$\pm$0.10  &  4.34$\pm$0.10  \\
Zn &  4.28$\pm$0.18  &  4.80$\pm$0.10  &  5.00$\pm$0.20  &  4.55$\pm$0.10  &  4.76$\pm$0.30  &  4.35$\pm$0.10  &  4.72$\pm$0.10  &  4.72$\pm$0.10  &  4.60$\pm$0.10  \\
Sr &  3.30$\pm$0.09  &  3.30$\pm$0.10  &  3.24$\pm$0.10  &  3.33$\pm$0.10  &  3.11$\pm$0.08  &  3.32$\pm$0.10  &  3.21$\pm$0.10  &  3.20$\pm$0.10  &  3.29$\pm$0.15  \\
Y  &  2.47$\pm$0.08  &  2.46$\pm$0.10  &  2.39$\pm$0.10  &  2.48$\pm$0.10  &  2.50$\pm$0.09  &  2.50$\pm$0.11  &  2.46$\pm$0.10  &  2.46$\pm$0.12  &  2.50$\pm$0.10  \\
Zr &  2.93$\pm$0.10  &  2.93$\pm$0.25  &  2.75$\pm$0.10  &  2.90$\pm$0.10  &  2.74$\pm$0.13  &  2.91$\pm$0.10  &  2.85$\pm$0.10  &  2.80$\pm$0.07  &  2.97$\pm$0.10  \\          
Ba &  2.60$\pm$0.06  &  2.93$\pm$0.10  &  2.35$\pm$0.09  &  2.63$\pm$0.08  &  2.63$\pm$0.17  &  2.52$\pm$0.10  &  2.54$\pm$0.10  &  2.58$\pm$0.14  &  2.55$\pm$0.08  \\

\hline
 
\end{tabular}
\end{center}
\begin{list}{}{}
\item[$^*$] Non-member. 
\end{list}
\end{table*}             

\begin{figure*}
\begin{center}

\includegraphics[width=16cm]{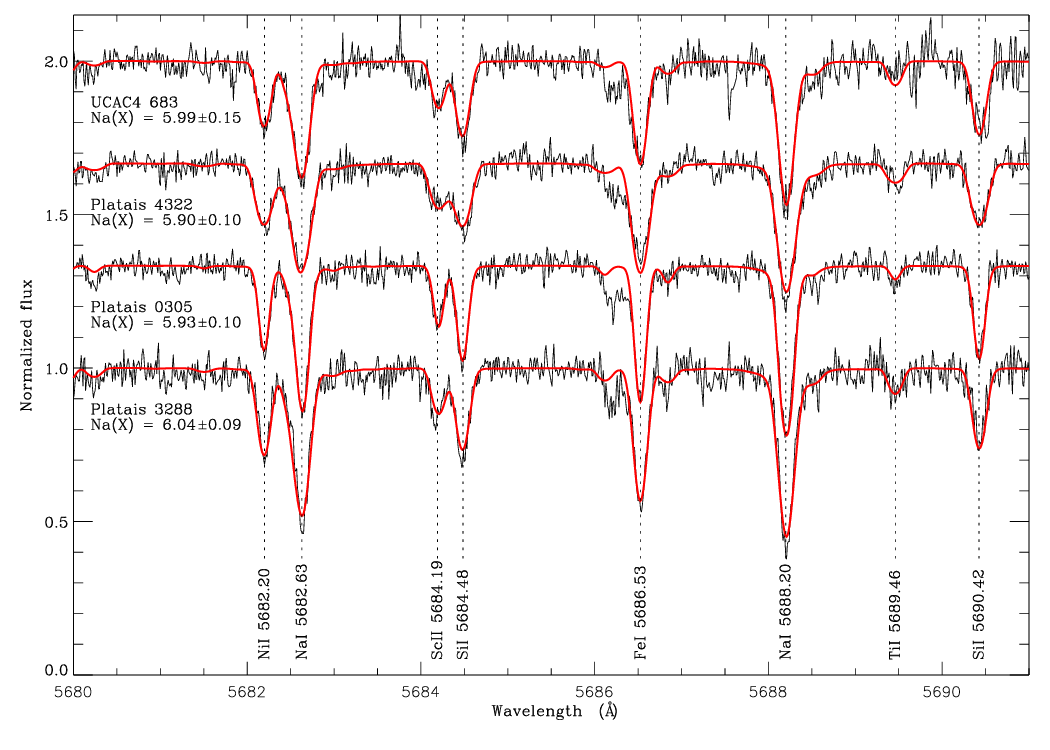}	
\caption{Example of spectral synthesis for the stars in our sample with lower $v \sin\,i$. The two most intense spectral features in this range (5680--5700\,\AA{}) correspond to \ion{Na}{I} lines, the abundance of which is shown.}
\label{spectra}
\end{center}
\end{figure*}

\end{document}